\documentclass[longauth]{aa}  
\usepackage{natbib}
\bibpunct{(}{)}{;}{a}{}{,} 
\usepackage{graphicx}
\usepackage{txfonts}
\usepackage{subfig}
\usepackage{soul}
\usepackage[dvipsnames]{xcolor}
\usepackage{amsmath}
\usepackage{hyperref}
\usepackage{orcidlink}
\hypersetup{
    colorlinks=true,
    citecolor=blue,
    linkcolor=blue,
    filecolor=magenta,      
    urlcolor=cyan}
\usepackage{caption}
\usepackage{lipsum}
\usepackage{animate}
\usepackage[normalem]{ulem}

\newcommand{\chandra}{{\it Chandra}}
\newcommand{\xmm}{{\it XMM-Newton}}
\newcommand{\nustar}{{\it NuSTAR}}

\begin{document} 

   \title{BASS}
   \subtitle{LV. Connecting X-ray variability with AGN physical properties and a new path to Cosmological distances}

\author{
Matilde Signorini\inst{1,2,15}\thanks{\email{matilde.signorini@esa.int}}\orcidlink{0000-0002-8177-6905},
Federica Ricci\inst{1,3,4}\orcidlink{0000-0001-5742-5980},
Alessia Tortosa\inst{3}\orcidlink{0000-0003-3450-6483},
Stefano Bianchi\inst{1}\orcidlink{0000-0002-4622-4240},
Fabio La Franca\inst{1}\orcidlink{0000-0002-1239-2721},
Franz E. Bauer\inst{5}\orcidlink{0000-0002-8686-8737},
Fiona A. Harrison\inst{10}\orcidlink{0000-0002-4226-8959},
Kohei Ichikawa\inst{7,8,9}\orcidlink{0000-0002-4377-903X},
Arghajit Jana\inst{6,11}\orcidlink{0000-0001-7500-5752},
Michael J. Koss\inst{12}\orcidlink{0000-0002-7998-9581},
Tingting Liu\inst{13},
Kyuseok Oh\inst{14}\orcidlink{0000-0002-5037-951X},
Alessandro Peca\inst{12,16}\orcidlink{0000-0003-2196-3298},
Meredith Powell\inst{17},
Claudio Ricci\inst{23,24,25}\orcidlink{0000-0001-5231-2645},
David B. Sanders\inst{18}\orcidlink{0000-0002-1233-9998},
Roberto Serafinelli\inst{6,3}\orcidlink{0000-0003-1200-5071},
Daniel Stern\inst{19}\orcidlink{0000-0003-2686-9241},
Benny Trakhtenbrot\inst{20}\orcidlink{0000-0002-3683-7297},
Ezequiel Treister\inst{21}\orcidlink{0000-0001-7568-6412},
Megan Urry\inst{16,22}\orcidlink{0000-0002-0745-9792}
}

\institute{
Dipartimento di Matematica e Fisica, Universit\`{a} degli Studi Roma Tre, via della Vasca Navale 84, I-00146 Roma, Italy
\and
INAF -- Osservatorio Astrofisico di Arcetri, Largo Enrico Fermi 5, I-50125 Firenze, Italy
\and
INAF -- Osservatorio Astronomico di Roma, Via Frascati 33, 00078 Monte Porzio Catone, Italy
\and
INAF -- Osservatorio di Astrofisica e Scienza dello Spazio di Bologna, Via Gobetti 93/3, I-40129 Bologna, Italy
\and
Pontificia Universidad Cat\'{o}lica de Chile, Instituto de Astrof\'{i}sica, Casilla 306, Santiago 22, Chile
\and
Instituto de Estudios Astrof\'{i}sicos, Facultad de Ingenier\'{i}a y Ciencias, Universidad Diego Portales, Av. Ej\'{e}rcito Libertador 441, Santiago, Chile
\and
Frontier Research Institute for Interdisciplinary Sciences, Tohoku University, Sendai 980-8578, Japan
\and
Astronomical Institute, Tohoku University, Aramaki, Aoba-ku, Sendai, Miyagi 980-8578, Japan
\and
Global Center for Science and Engineering, Faculty of Science and Engineering, Waseda University, 3-4-1 Okubo, Shinjuku, Tokyo 169-8555, Japan
\and
Cahill Center for Astronomy and Astrophysics, California Institute of Technology, Pasadena, CA 91125, USA
\and
Department of Physics, SRM University--AP, Amaravati 522240, India
\and
Eureka Scientific, 2452 Delmer Street, Suite 100, Oakland, CA 94602-3017, USA
\and
Department of Physics and Astronomy, Georgia State University, 25 Park Place, Suite 605, Atlanta, GA 30303, USA
\and
Korea Astronomy and Space Science Institute, 776 Daedeokdae-ro, Yuseong-gu, Daejeon 34055, Republic of Korea
\and
European Space Agency (ESA), European Space Research and Technology Centre (ESTEC), Keplerlaan 1, 2201 AZ Noordwijk, The Netherlands
\and
Department of Physics, Yale University, P.O. Box 208120, New Haven, CT 06520, USA
\and
Leibniz-Institut f\"{u}r Astrophysik Potsdam (AIP), An der Sternwarte 16, 14482 Potsdam, Germany
\and
Institute for Astronomy, University of Hawai`i, 2680 Woodlawn Drive, Honolulu, HI 96822, USA
\and
Jet Propulsion Laboratory, California Institute of Technology, 4800 Oak Grove Drive, MS 169-224, Pasadena, CA 91109, USA
\and
School of Physics and Astronomy, Tel Aviv University, Tel Aviv 69978, Israel
\and
Instituto de Alta Investigaci\'{o}n, Universidad de Tarapac\'{a}, Casilla 7D, Arica, Chile
\and
Yale Center for Astronomy \& Astrophysics and Department of Physics, Yale University, P.O. Box 208120, New Haven, CT 06520-8120, USA
\and
Department of Astronomy, University of Geneva, ch. d'Ecogia 16, 1290, Versoix, Switzerland 
\and
Instituto de Estudios Astrof\'isicos, Facultad de Ingenier\'ia y Ciencias, Universidad Diego Portales, Av. Ej\'ercito Libertador 441, Santiago, Chile 
\and
Kavli Institute for Astronomy and Astrophysics, Peking University, Beijing 100871, People's Republic of China
}

\titlerunning{BASS LV}
\authorrunning{M. Signorini et al.}

   \date{\today}

  \abstract
{X-ray variability is a well-established characteristic of active galactic nuclei (AGN), known to correlate inversely with both the supermassive black hole mass (\( M_{\rm BH} \)) and luminosity, although the degree of each remains a topic of debate. The potential of X-ray variability as a proxy for $M_{\rm BH}$ or for intrinsic $L_{\rm X}$ has led to proposals to use AGN as standard candles to test cosmological models. However, the large intrinsic dispersion in these relations has limited their practical applications.
In this work, we investigate the dependence of X-ray variability on AGN physical properties using a sample of 134 Seyfert 1 galaxies from the BAT AGN Spectroscopic Survey (BASS), which is the largest sample to date, more than three times larger than those used in previous studies. Contrary to earlier findings, we observe that X-ray variability correlates with luminosity just as strongly as with $M_{\rm BH}$. Furthermore, we still do not find evidence for the expected anti-correlation between variability and Eddington ratio,  even when using refined bolometric luminosities from SED fitting to compute the Eddington ratio. From a cosmological perspective, the increased sample size reduces the scatter in the \(\log L\)--\(\log \sigma^2_{\rm NXS}\) relation to \(\sim 0.63\)~dex—a significant improvement over previous results, but still too large to serve as competitive standard candles, when compared to SNIa (uncertainties on distances of $\sim$5-10\%) or the $L_X-L_{UV}$ relation in quasars (uncertainties of 10-12\%). We tested including the width of broad emission lines as additional parameters, but found that this does not significantly lower the observed dispersion, contrary to previous studies on smaller samples. Finally, we discuss how future X-ray missions such as \textit{AXIS} and \textit{NewAthena} will improve this scenario by enabling precise variability measurements for thousands of AGN up to redshift \( z \sim 3 \), thereby enabling it as a new cosmological probe.}
\keywords{galaxies: active; quasars: general; quasars: supermassive black holes; methods: statistical}
\maketitle
\section{Introduction}
\label{intro}
Active Galactic Nuclei (AGN) are powered by accretion onto supermassive black holes (SMBHs) and are among the most luminous sources in the Universe. They are observed across the entire cosmic time, up to redshift $\sim$10 \citep{Mortlock11, Wang21, Napolitano25}, and they play a key role in the formation and evolution of galaxies and black holes \citep{Fabian12, Kormendy13}.
One of the defining features of AGN is their variability, which is observed across entire electromagnetic wavebands, from radio to X-rays \citep[e.g.,][]{Cristiani96,Ulrich97, McHardy2004}. This variability occurs on a range of timescales, from hours to decades to even larger ones, and provides a unique tool for probing the physical processes both close to and far from the central SMBH. In the X-ray band, variability is observed from very short (less than 10$^3$~s) to long (years) timescales \citep{Uttley05, McHardy2004, Peca25, Sartori18}. It is thought to originate from processes within the innermost regions of the accretion disk and corona, making it a powerful diagnostic of the AGN structure and dynamics \citep{Mushotzky93, Ulrich97, Uttley14, Middei17, Cackett21, DeMarco22}.\\
A well-established correlation exists between AGN X-ray variability amplitude and both the black hole mass ($M_{\rm BH}$) and the luminosity, with more massive, more luminous AGN showing systematically lower variability at different wavelengths \citep{Nandra97, Turner99, Lu01, O'Neill05, McHardy06, Gierlinski08, Zhou10, Ponti12, Kelly13, Tortosa23}. It has been suggested that the true physical driver of the variability is the $M_{\rm BH}$, with the relation with the luminosity emerging as a consequence, due to the dependence of luminosity on the black hole mass itself \citep[e.g.,][]{ONeill05, Papadakis04, Ponti12, Tortosa23}. Recent hard–X results further show that the amplitude of the variability, measured e.g. by the excess variance \(\sigma^2_{\mathrm{NXS}}\), primarily anti-correlates with \(M_{\rm BH}\), with no significant dependence on \(\lambda_{\rm Edd}\) on \(\sim10\) ks timescales \citep{Serafinelli24}. At the same time, many studies of X-ray AGN variability use small datasets, and the dependence of variability on $M_{BH}$, luminosity, and accretion efficiency remains poorly understood.\\
Recent large-sample studies further highlighted the importance of variability scaling relations at different wavelengths. \cite{Georgakakis26} used SRG/eROSITA multi-epoch X-ray photometry to measure the  ensemble excess variance for nearly ten thousand QSOs, extending the $\sigma^2_{\rm NXS}$–$M_{\rm BH}$ anti-correlation to the highest black hole masses and finding non-monotonic ('U-shaped') dependence on the Eddington ratio. \cite{Helias26} analysed Gaia, SDSS and ZTF light curves for a large AGN sample and confirmed the strong link between variability timescales and black hole mass in the optical wavelengths. 
%a

The fact that AGN are observed over the entire cosmic history makes them valuable not only for understanding SMBH growth and galaxy evolution but also as potential cosmological probes. Their brightness and ubiquity make them ideal candidates for tracing the expansion history of the Universe and testing cosmological models at both early and late epochs. Accurate distance measurements across a wide range of redshifts are essential for constraining the nature of dark energy and the geometry of the Universe \citep[e.g., ][]{Riess98, Perlmutter99, Planck18}. While Type Ia supernovae have been the primary standard candles up to $z \sim 2$, their limited redshift reach motivates the search for complementary probes capable of extending distance measurements to the epoch of reionisation. In this context, AGN offer a unique opportunity, as they are among the brightest persistent sources in the Universe.
Various empirical correlations involving AGN luminosity have been explored to investigate the possible use of AGN as standard candles. Examples include the Baldwin effect—linking the continuum luminosity to the equivalent width of broad emission lines \citep{Baldwin77}—and the radius–luminosity (\( R\)--\(L \)) relation measured via reverberation mapping (RM) of the broad-line region \citep[BLR; e.g.,][]{Watson11, Khadka22}. While promising in concept, these correlations typically suffer from large intrinsic dispersions (up to $\sim$0.8~dex), which limit their utility for precise distance determinations, or they are applicable over restricted redshift ranges. Another proposed approach uses water megamaser emission to determine geometric distances \citep{Pesce20}, but this method is constrained by the rarity of such sources and the high observational demands required for accurate measurements.
A more recent and promising method exploits the tight empirical relation between the X-ray and UV luminosities in quasars \citep{Lusso20}. Thanks to careful sample selection and correction for systematic biases, the observed dispersion has been reduced from initial values of $\sim$0.40~dex \citep{Lusso16} to as low as $\sim$0.12~dex in the best samples \citep{Sacchi22}, with an estimated intrinsic scatter below 0.1~dex \citep{Signorini24}.

The anti-correlation between variability (specifically measured as the X-ray excess variance) and luminosity suggests that variability can also serve as a proxy for intrinsic luminosity \citep{Ponti12, LaFranca14}. However, in this case, previous work showed a high scatter in the relation, making the derived distances not particularly useful for cosmological applications. \citet{LaFranca14} studied a sample of 40 radio-quiet, X-ray unobscured AGN, and found that including broad line information in the analysis significantly lowered the dispersion (see Section \ref{sec: distances}).  In said work, including information from the FWHM decreased the dispersion values from 1.32~dex to 0.93~dex when using H$\beta$ and from 1.28~dex to 0.56~dex when using Pa$\beta$. 

Overall, the study of X-ray variability in AGN is key not only to understanding the physical mechanisms governing the X-ray corona and its connection with the accretion process, but also to exploring its potential as a cosmological tool. In this work, we use the Burst Alert Telescope (BAT) AGN Spectroscopic Survey \citep[BASS;] []{Koss2022a} to systematically investigate X-ray variability in a sample of $\sim$150 objects with multiwavelength information, significantly increasing statistics and completeness compared to previous studies. Our goals are: \begin{itemize}
    \item to investigate the dependence of X-ray variability on physical AGN parameters such as $M_{\rm BH}$ and the Eddington ratio ($\lambda_{\rm Edd}$), taking advantage of the recently published SED-based estimates of $\lambda_{\rm Edd}$ \citep{Gupta24} and the extensive multi-wavelength characterisation of BASS; 
    \item to assess the reliability of using X-ray variability as a proxy for intrinsic luminosity and as a potential distance indicator, quantifying how the statistical power of the BASS sample affects the scatter in the excess variance–luminosity relation;
    \item to test whether incorporating broad-line width information, following \citet{LaFranca14}, can further reduce the observed dispersion;
    \item to evaluate how future X-ray facilities will enhance the use of AGN variability for cosmological distance measurements.   
\end{itemize} 
The paper is structured as follows: in Section  \ref{sec:sample} we describe the sample and its properties; in Section \ref{sec: fitting} we describe how we performed the fits throughout the paper; in Section \ref{sec: exvarother} we study the dependence of X-ray variability on the AGN properties, in particular the $M_{\rm BH}$, the monochromatic luminosities at different wavelengths, and the $\lambda_\mathrm{Edd}$; in Section \ref{sec: distances} we test the use of this sample as a calibrator for AGN distance measurement; in Section \ref{sec: future} we discuss future implementations of the method in light of upcoming new observatories; in Section \ref{sec: conclusions} we draw our conclusions. Throughout this paper, we consider a flat $\Lambda$CDM cosmology with $H_0 = 70$ km s$^{-1}$ Mpc$^{-1}$, $\Omega_{\Lambda}$ = 0.73, and $\Omega_{m}$ = 0.27, as in \cite{Tortosa23} from which we derive the luminosities used in this study. Correlations are considered significant when they yield p-values lower than 0.01, corresponding to $\sim$2.6 $\sigma$ under the null hypothesis of non-correlation, assuming a Normal distribution. Errors are reported at the 68\% confidence level or as the 16$^{th}$ and 84$^{th}$ percentiles, unless otherwise stated. Regarding the measurements of X-ray excess variance (see Section \ref{subsec: exvar}), we consider upper limits on the variability measurements that are consistent with zero at one sigma. 

\section{Sample selection and properties}
\label{sec:sample}
The starting point of our work is the BASS sample\footnote{https://www.bass-survey.com/}, constructed from X-ray–detected sources identified by the all-sky survey carried out with the Burst Alert Telescope \citep[BAT;][]{Barthelmy05} aboard the Neil Gehrels Swift Observatory \citep{Gehrels04}. BASS is a crucial survey for the local AGN census, as it probes AGN up to Compton-thick obscuration levels (which is, with a column density of $N_H>10^{24}cm^{-2}$; \citealp{Ricci15, Koss16}), providing a comprehensive view of accreting supermassive black holes in the nearby Universe. The Swift/BAT data \citep[14-195 keV]{Baumgartner13, Oh18} is complemented with observations in the softer X-rays (0.5-10 keV) from \chandra, \xmm, and \nustar~\citep[e.g.,][]{Ricci_C_17, Peca25}. The survey also includes multi-wavelength follow-up data, from optical emission \citep{Oh22}, to near-infrared \citep[][Gillette et al., in prep.]{Ricci22, denBrok2022}, to mid- and far-IR emission observed by WISE, IRAS, Spitzer, AKARI, and Herschel \citep{Ichikawa17, Ichikawa19, Shimizu17}. Further insights are provided by mm/radio emission studies \citep{Koss21, Kawamuro22, RicciC23}, covering the broadest possible spectral range.\\
The first BASS data release \citep[DR1;][]{Koss17} reported $M_{\rm BH}$ and X-ray properties for all 838 AGN in the Swift/BAT 70-month catalog \citep{Ricci_C_17}. The second data release \citep[DR2;][]{Koss2022a} refined these results, providing more reliable and uniformly assessed $M_{\rm BH}$ values for 780 unbeamed AGN in the same catalog. $M_{\rm BH}$ were derived from broad Balmer lines and/or the RM technique for type 1 AGN, and using maser, dynamical modeling, or velocity dispersion measurements for type 2 AGN \citep{Koss22c, Mejia-Restrepo2022, Ricci_F_2022}. DR2 also calculated $\lambda_{\rm Edd}$ ($L_{\rm bol}/L_{\rm Edd}$), using bolometric luminosities derived from intrinsic 14–150 keV luminosities, applying a bolometric correction factor of 8 \citep{Ricci_C_17, Koss2022a}.

In this work we started from the sample described in \cite{Tortosa23}, which consists of 151 radio-quiet unobscured ($N_{\rm H}$/cm$^{-2}$ < 10$^{22}$) type 1 AGN from the BASS sample that had archival \xmm observations and whose excess-variance has been measured by \cite{Tortosa23} (see sect. 2.1.). In addition to the X-ray variability data, we cross-matched the 151 objects with those that have optical and/or near-infrared broad-line spectral measurements from previous BASS collaboration works. This brings the sample to 144 objects with at least one measured Hydrogen broad line. For all these objects, single-epoch $M_{\rm BH}$ measurements are available, with an average uncertainty of $\sim$0.5 dex \citep{Koss2022a}. 
Recent analysis of the BASS sample regards the presence of Double Peak Emission objects \citep{Ward25}, which could be due to the emission of rotating gas in the accretion disk. The presence of these components means the broad-line widths are not directly tied to the rotational velocity of the BLR, introducing bias in virial SMBH mass estimates based on broad-line widths. Among the 144 objects with X-ray excess variance measurement and broad line measurements, 10 show strong signs of being DPEs. Therefore, we removed them from the sample, as their single-epoch $M_{\rm BH}$ could be strongly affected. The final sample size is of 134 objects. Among those, 32 also have $M_{\rm BH}$ estimates from RM campaigns (the `Reverberation mapping subsample'). 

\begin{figure*}
	\centering
    
\includegraphics[width=0.47\linewidth]{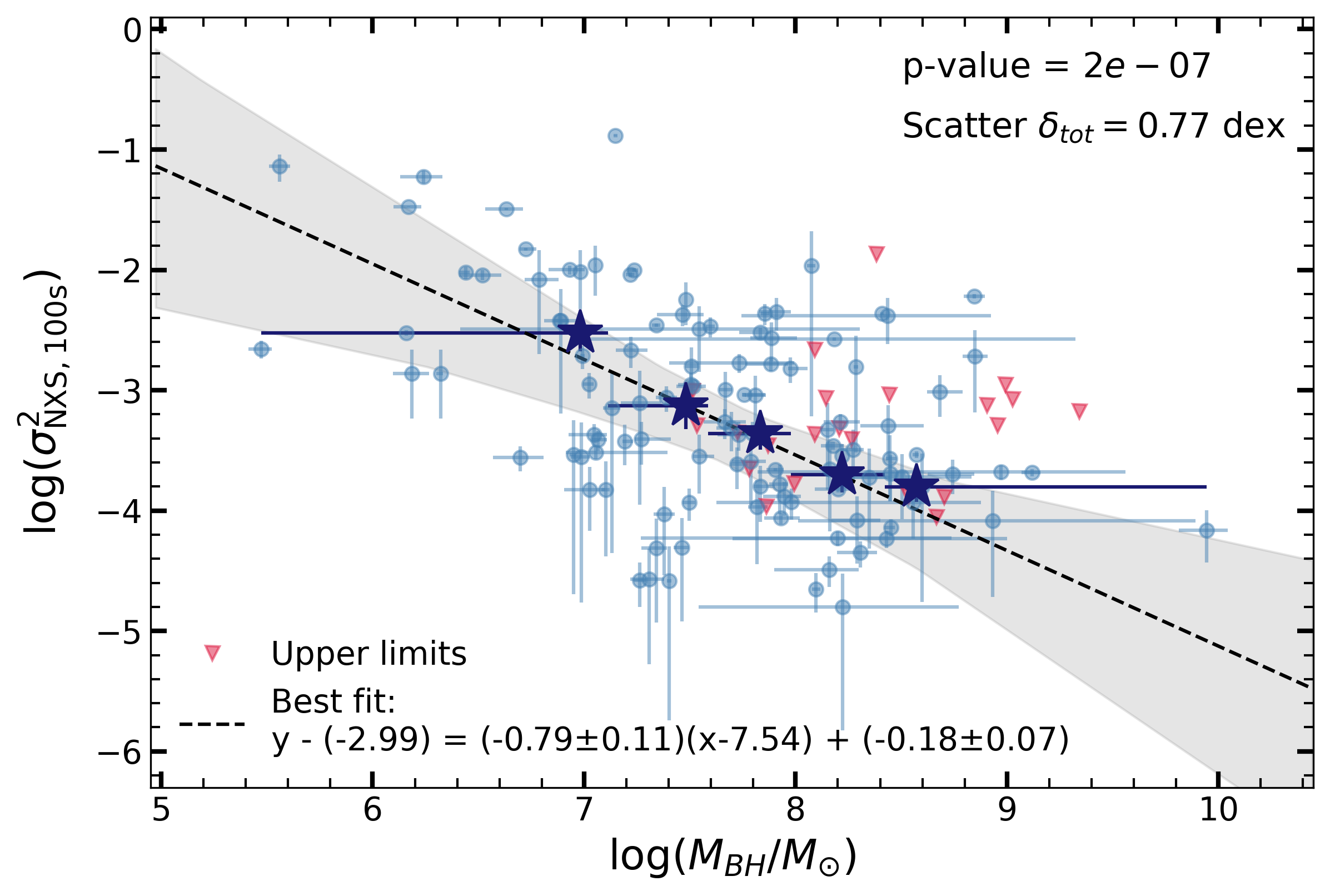}
\includegraphics[width=0.47\linewidth]{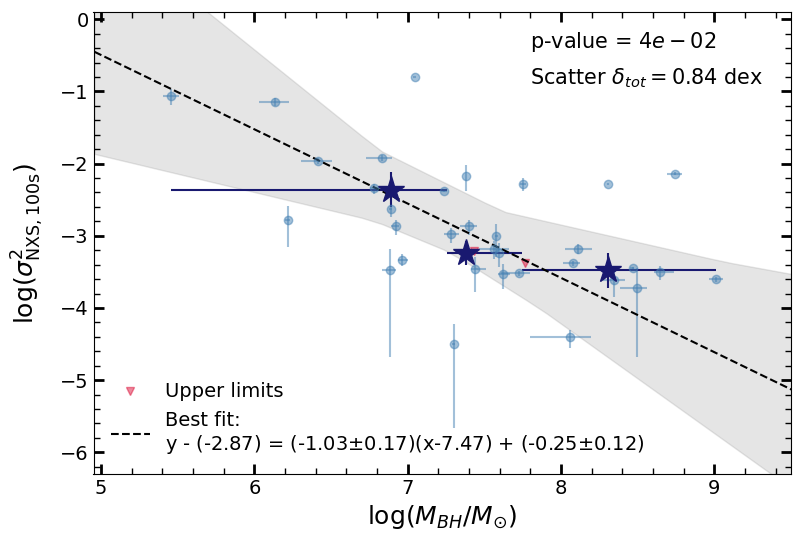}
\caption{Left: Relation between the 2-10 keV excess variance ($\sigma_{\rm NXS}^2$) (obtained with time bins of 100~s) and the black hole mass ($M_{BH}$) for the sample of 134 objects. Filled cyan circles show full detections; red down-pointing triangles show upper limits; the grey shaded area shows the confidence interval at 1$\sigma$ for the best fit log-linear relation. Blue stars represent the median value of the $M_{BH}$ and the excess variance in six $M_{BH}$ bins, chosen so that in each bin there are at least 20 objects. Blue error bars indicate the bin width. Blue stars are placed solely to guide the eye, as the fit is performed on the entire sample. The p-value of the relation and the total scatter $\delta_{\rm tot}$ are displayed, together with the best fit result. The best fit is reported on the plot as $y-\bar{y} = a(x -\bar{x}) + b'$, as the fit is performed by centering the variables at their average values $\bar{x}$ and $\bar{y}$. Therefore, the true intercept of the relation is to be recovered as $b=b'+\bar{y}-a\bar{x}$. Right: same but for the Reverberation Mapping sample. The relation is steeper but consistent with what is found for the whole sample.}
	\label{fig:exvar_Mbh}
\end{figure*}

\subsection{Excess variance}
\label{subsec: exvar}
The excess variance is a measure of the variability amplitude in a light curve. It is calculated as the difference between the total variance of a light curve and the mean squared error, normalised by the square of the mean flux across \(N\) time bins \citep[][]{Nandra97, Turner99}. Here, \(N\) represents the number of time bins within a given lightcurve, while \(x_i\) and \(\sigma_i\) are the flux and its associated error in each interval, respectively. The excess variance is formally defined \citep{Vaughan03} as follows:
\begin{equation}
\sigma_\mathrm{NXS}^2 = \frac{S^2 - \sigma^2}{\bar{x^2}}
\end{equation}
where \(S^2\) is the sample variance and \(\sigma^2\) is the mean squared error. The sample variance, \(S^2\), corresponds to the integral of the power spectral density (PSD) $P(\nu)$ over a given frequency range (\(\nu_1\) to \(\nu_2\)), representing the contribution to the variance from variations on timescales between \(1/\nu_1\) and \(1/\nu_2\):

\begin{equation}
    \langle S^2 \rangle = \int_{\nu_{1}}^{\nu_{2}}P(\nu) d\nu \, .
\end{equation}

Although \(\sigma_\mathrm{NXS}^2\) is a reliable estimator of the intrinsic variance of a source, it is not free from biases. Its value depends on the integral of the PSD over specific frequencies and is influenced by several factors, including the duration of the monitoring interval, the red-noise nature of X-ray variability, and the effects of cosmological time dilation. The latter introduces dependencies on the source's redshift \citep{Lawrence93, Green93, Papadakis08, Vagnetti11, Vagnetti16}. In our sample of 134 type 1 AGN, which consists of local AGN (\(z_\mathrm{med} = 0.035\)), the impact of redshift on excess variance is negligible. However, we must account for potential biases related to the varying exposure times of our observations and the red-noise character of the light curves to ensure the robustness of our results.

\cite{Tortosa23} measured the excess variance for 153 unobscured AGN from the BASS sample observed with \xmm, using the EPIC-pn \citep{Struder01} light curves, extracted using both 100~s and 1000~s binning in the 0.2--10~keV energy band, and 100~s in the 0.2--1 keV(soft), 1--3~keV (medium), and 3--10~keV (hard) energy bands (see Sect~2.2 of \citealt{Tortosa23} for more details). They selected observations that had cleaned exposure times larger than 10~ks and which had at least 10 counts in those 10~ks chunks and in each (rest frame) energy band used in the analysis, i.e., 0.2--1, 1--3, and 3--10~keV, and for each time bin of 100~s and 1000~s. This was done to avoid having too few counts in the 10~ks independent light curve to constrain the excess variance. A total of 151 sources ($\sim 500$ observations) fulfill these criteria. Of these 151 sources, 46 have upper limit measurements for the excess variance when using a 100~s bin. An object is considered an upper limit when the one-sigma uncertainty on the excess variance is equal or higher than the measured value. If an object had cleaned exposure longer than 10 ks, the light curve was split into as many 10 ks segments as possible; the excess variance was computed separately in each segment (with \(N\) time bins in each segment) and the per-segment values were then combined per source using the median. This uniform segmentation mitigates biases due to heterogeneous monitoring lengths across sources.
Because the excess variance represents the integral of the PSD over the sampled frequency range, it provides a reliable proxy for variability amplitude only when that range lies entirely on one side of the PSD break. In other words, it must sample a region of the PSD with a single slope (either the low-frequency or the high-frequency part of the broken power-law). Adopting the mass–accretion scaling of the PSD break (e.g. \citealt{McHardy06}), the break timescale increases with black hole mass and decreases with the Eddington ratio ($T_{\rm B}\!\propto\! M_{\rm BH}^{\alpha}\lambda_{\rm Edd}^{-1}$, with $\alpha\!\approx\!1$). For the BASS AGN in our sample ($M_{\rm BH}\!\gtrsim\!10^{6}\,M_\odot$) and moderate $\lambda_{\rm Edd}$, the break typically occurs on timescales longer than $\sim10$–$20\,\mathrm{ks}$ (see also \citealt{Ludlam2015}), so we are confident that we are always looking at the same PSD regime.

Regarding other possible source of biases, we note how using a fixed‑length 10 ks segments for all objects ensures that the same PSD frequency range is sampled for each source, therefore mitigating the red‑noise and exposure‑time biases. Furthermore, \cite{Tortosa23} find results that are consistent when using a 100~s and a 1000~s binning. Since changing the bin size by a factor of 10 shifts the sampled frequency range by a factor of 10, analogous results indicate that neither high‑frequency red‑noise nor Poisson noise dominates the excess variance estimates.
\\
The presence of upper limits makes it more difficult to constrain the dependence of variability on other physical properties. To start our work we looked for a way to possibly reduce the upper limits fraction with the already available data. As all of the 134 objects in our sample had at least 20~ks observations, we derived the excess variance values using extended chunks of 20~ks instead of 10~ks, as a longer light curve should be more capable to constrain variability, especially for objects with higher black hole masses, that are less variable. This choice resulted indeed in a significant decrease in the upper limit fraction with respect to the work presented in \citet{Tortosa23}. Among the 134 objects of our sample, the upper limits range from 46 to 22, which is, from 34\% to 16\% of the total samples. This shows how monitoring for longer times is essential for reliable variability estimates, at least for $M_{\rm BH}>10^6$~M$_\odot$. 
As in \cite{Tortosa23} the excess variance values are calculated both using time bins of 100~s and 1000~s. The following results are all obtained with the 100~s binning values, but we also tested and confirmed that analogous results are obtained using time bins of 1000~s.

\subsection{Emission line data}
In our sample, all of the 134 objects have broad H$\alpha$ and H$\beta$ measurements from \cite{Mejia-Restrepo2022}; 67 objects have Paschen lines measurements. 
For the broad lines H$\alpha$, H$\beta$, Pa$\alpha$, Pa$\beta$, we have estimates of the flux and FWHM from spectral fitting. For the data coming from the previous BASS releases \citep{Lamperti2017, denBrok2022, Mejia-Restrepo2022, Ricci_F_2022}, we refer the reader to those papers for details about the spectral fitting. Regarding the new DR3 objects, a fitting approach similar to those previous works has been adopted. More details will be available in the upcoming paper (Gillette et al., in prep.). Given the results from the spectral fitting, we only included in the analysis those objects for which the flux and FWHM could be fully constrained (i.e., we removed upper limits), and we considered only objects flagged as having a satisfactory fit. This leaves us with 115 objects with H$\alpha$, 88 objects with H$\beta$, 46 objects with Pa$\alpha$, and 50 objects with Pa$\beta$ measurements.

\subsection{Eddington ratio}
It is our interest to study the relation between excess variance and Eddington ratio (see Section \ref{sec: exvarother}). In this work, we take advantage of the $\lambda_\mathrm{Edd}$ published by \cite{Gupta24}, where Spectral Energy Distribution (SED) fitting is used to derive the bolometric luminosity, instead of using a bolometric correction on monochromatic luminosity. Thanks to the extensive multi-wavelength data of the BASS sample, it was possible to obtain a bolometric luminosity from SED fitting for 115 out of the 134 objects in our sample. The missing objects are those for which the UVOT monitoring was insufficient, so the disk model could not be constrained \citep{Gupta24}. As the SED fitting takes into account the multi-wavelength information, instead of the correction on a monochromatic luminosity, we argue that these values can be considered to be more reliable in the final estimate of the $\lambda_\mathrm{Edd}$ of our sources. The second ingredient to derive the $\lambda_\mathrm{Edd}$ is the $M_{\rm BH}$. For our sample, this estimate is obtained with single-epoch measurements using the H$\beta$ line for 102 objects, while 32 have $M_{\rm BH}$ estimates coming from reverberation mapping. In cases where reverberation mapping values were available, we prioritised their use due to their higher reliability. Overall, the objects in our sample span in $\lambda_{\rm Edd}$ from $\sim$0.005 to $\sim$1.

\begin{figure}
	\centering
\includegraphics[width=0.85\linewidth]{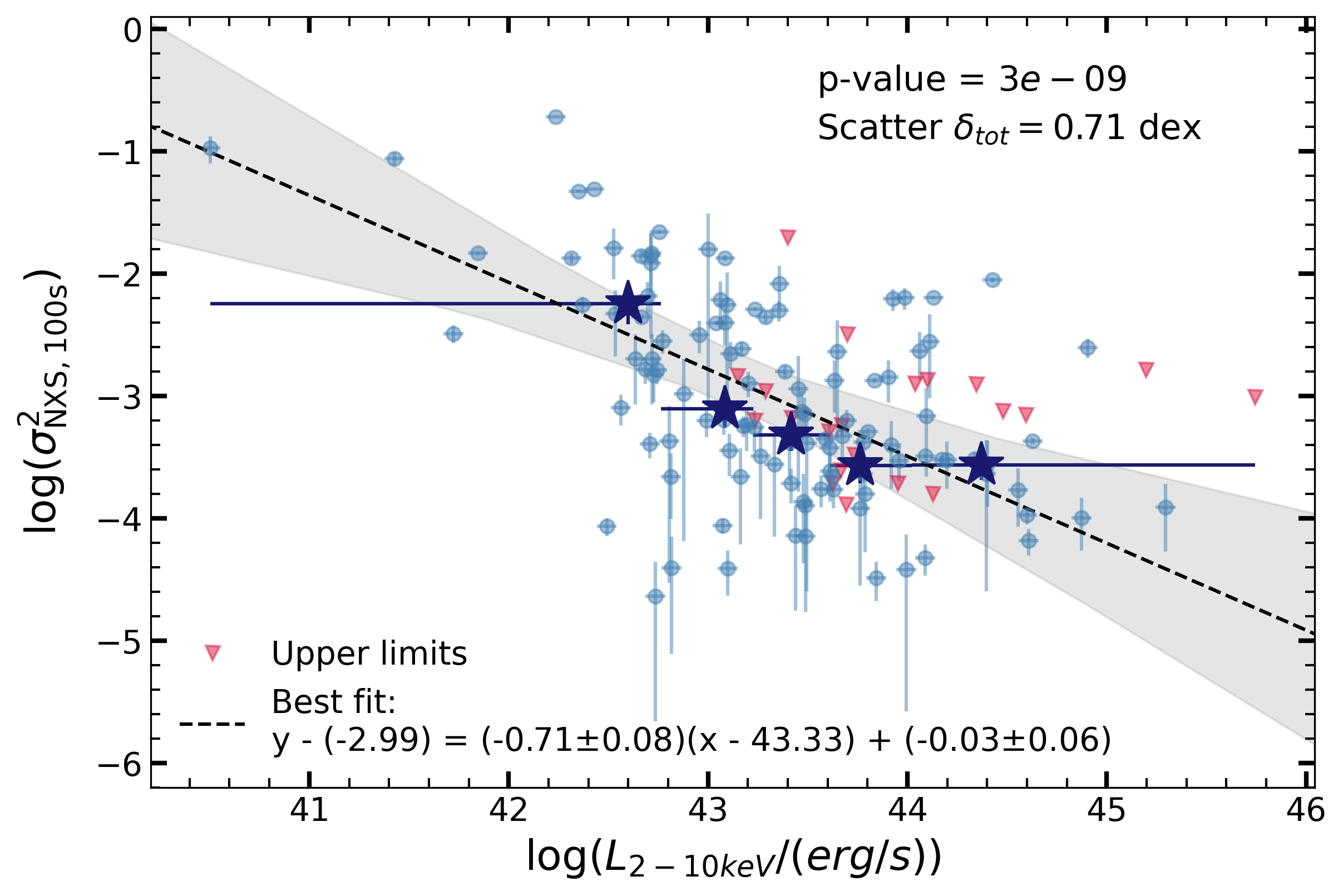}
\includegraphics[width=0.85\linewidth]{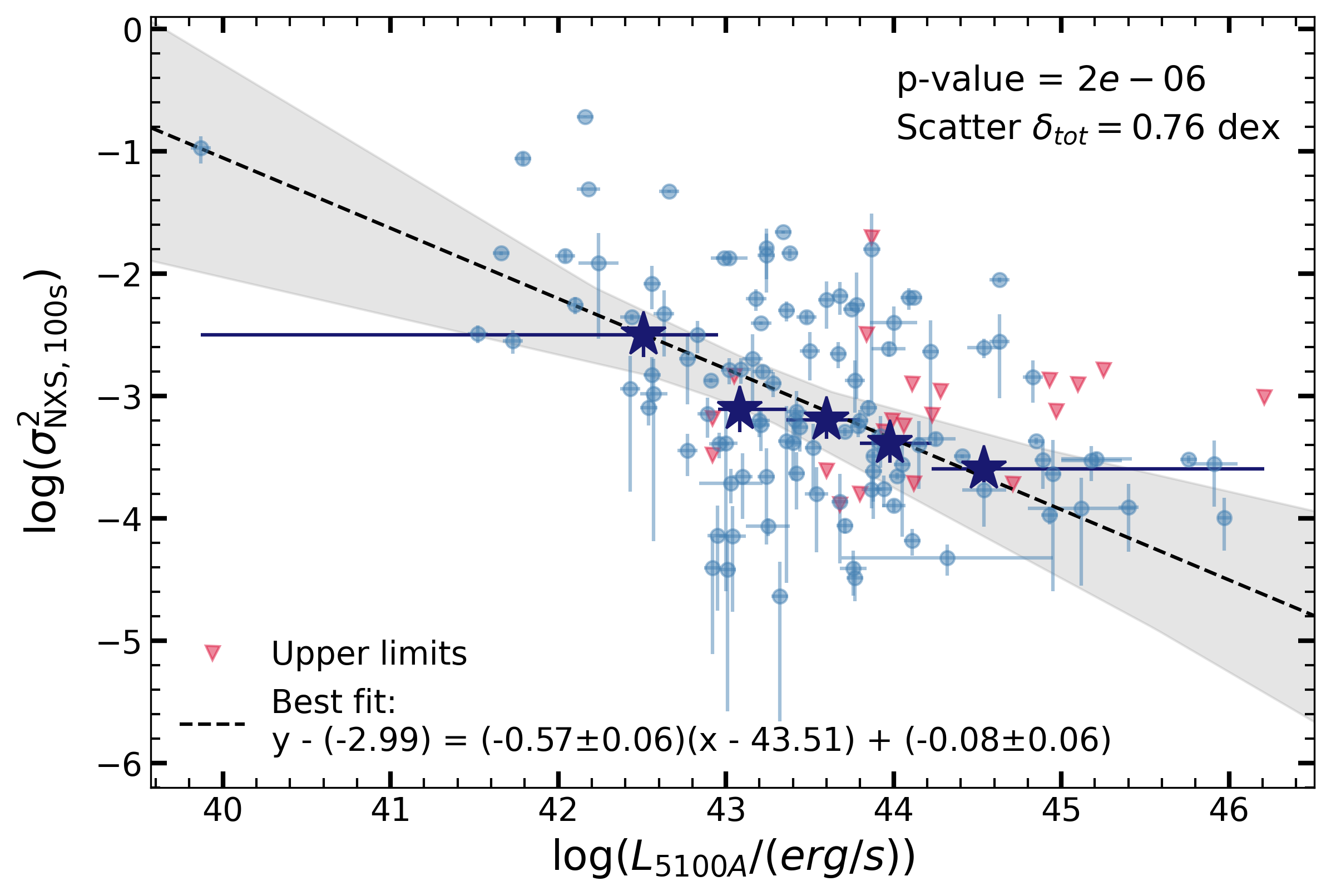}
\includegraphics[width=0.85\linewidth]{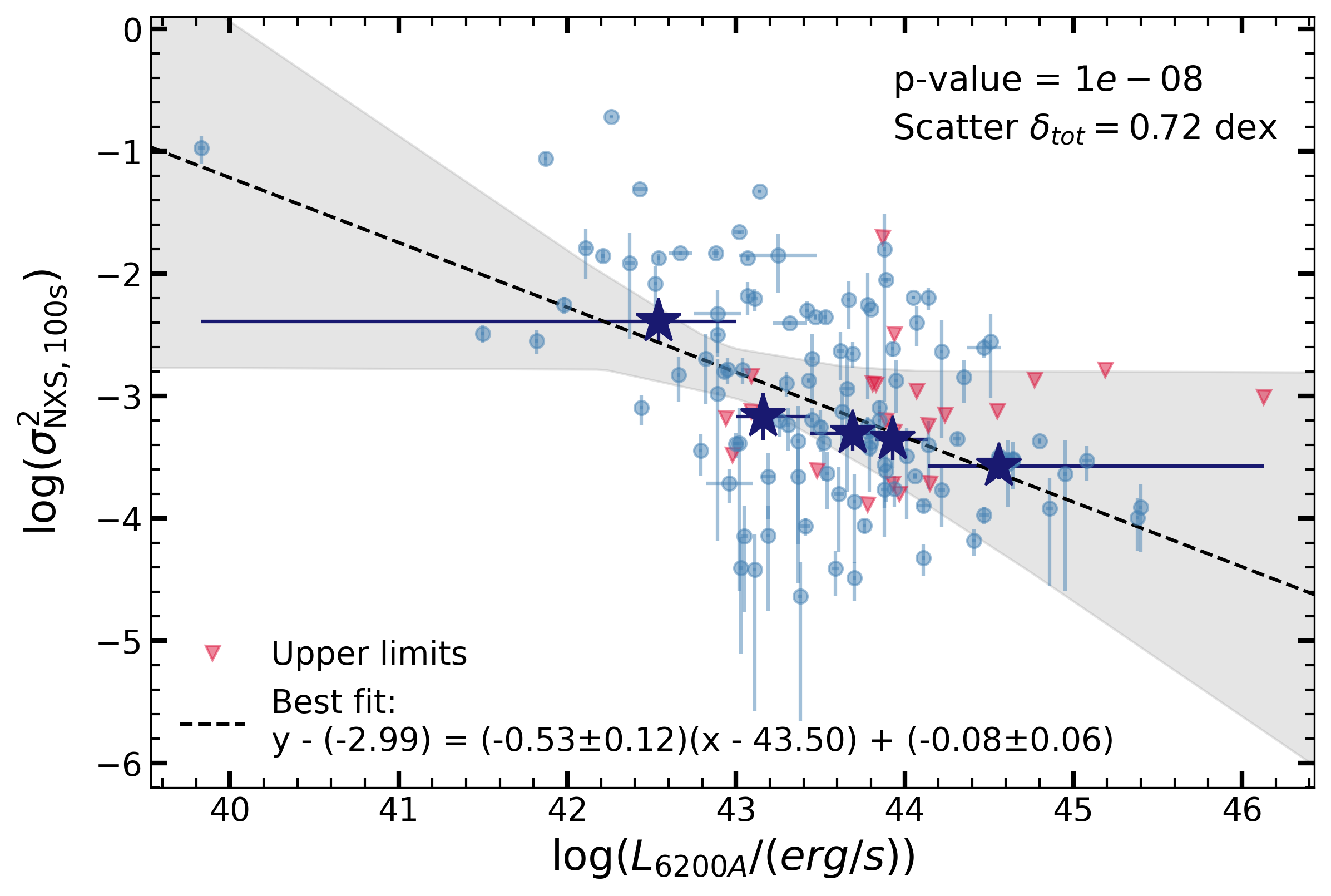}

\caption{Relation between the 2-10~keV excess variance (obtained with time bins of 100~s) and different monochromatic luminosities. The legend is the same as in Figure \ref{fig:exvar_Mbh}. Upper plot: 2-10~keV monochromatic luminosity; middle plot: 5100\AA{} monochromatic luminosity; lower plot: 6200\AA{} monochromatic luminosity. Sample size N=134.}
	\label{fig:exvar_L}
\end{figure}

\section{Data fitting}
\label{sec: fitting}
Throughout this study, we aim to fit log-linear relations between the X-ray excess variance and other physical quantities of the AGN in our sample. The fitting process is complicated by several challenges: (i) the presence of upper limits, (ii) significant uncertainties on both the dependent and independent variables, and (iii) asymmetric uncertainties. Importantly, our measurement uncertainties are first estimated in linear units; when transformed to logarithmic space, the upper and lower errors become asymmetric. Consequently, simply averaging upper and lower errors to define the weights would bias the fit; instead, we adopt an iterative scheme that assigns the appropriate upper or lower uncertainty to each point relative to a provisional best-fit line, and we validate this approach with mock datasets (Appendix~\ref{app: method}).
To ensure the reliability of our fitting method, we validated it extensively using mock datasets specifically designed to mimic the statistical properties of our sample (Appendix \ref{app: method} for details). These mock tests allowed us to evaluate the method's ability to recover the true underlying relations, meaning the values of the slope and the intercept, under varying fractions of upper limits and uncertainty configurations.
Our final fitting approach is iterative. Initially, we perform a simple fit using the \texttt{lmfit} Python library \citep{lmfit} with a linear model $y = a\cdot x+b$, considering as weight for each point the average uncertainty on both x and y, calculated as the mean of the upper and lower uncertainties. This preliminary fit provides a temporary best-fit line. Next, we classify each data point as lying above or below this line and, based on this classification, assign the upper or lower uncertainties to both the x and y axes. This process is repeated iteratively, with the fit recalculated at each step, until the best-fit parameters converge to a stable solution. Uncertainties are then derived on the best-fit values of the slope and the intercept with a bootstrap procedure. We validated this iterative method using mock datasets and found it to be robust for cases where the fraction of upper limits is below 30\%. Given that our sample includes only 16\% upper limits, we are confident on the reliability of our results. We also always performed the fit, centering the data, which is subtracting the mean values on x and y to minimise the covariance between the obtained slope and intercept of the linear fit. The mean values are then added to the best fit value of the intercept \textit{b} to recover the true value. 

\begin{table*}[ht]
\caption{Summary of fitting results for different relations}

\centering
\resizebox{\linewidth}{!}{%
\begin{tabular}{lccccc}
\hline
Relation & $a$ & $b$ & $\delta_\mathrm{tot}$ (dex)& $\delta_\mathrm{err}$ (dex)& $\delta_\mathrm{int}$ (dex) \\
\hline
%$\sigma^2$ vs. $M_\mathrm{BH}$ (all)       & $-0.79\pm 0.10$  & $2.76\pm 0.06 & 0.78$ & 0.25 & 0.73  \\
$\sigma^2$ vs. $M_\mathrm{BH}$ & $-0.78\pm 0.11$  & $2.76\pm0.07$   &  0.77  & 0.26   & 0.65 - 0.73   \\
$\sigma^2$ vs. $L_{2-10\,\rm keV}$ & $-0.71\pm0.08$  &  $27.64\pm0.06$ & 0.73  &  0.23 & 0.71  \\
$\sigma^2$ vs. $L_{5100 \AA}$  & $-0.57\pm0.06$  & $21.73\pm0.06$  & 0.76  & 0.23  &  0.74 \\
$\sigma^2$ vs. $L_{6200 \AA}$  & $-0.53\pm0.15$  &  $20.02\pm0.06$ & 0.73  & 0.28  & 0.67  \\

\hline
\end{tabular}
}
\tablefoot{The columns represent the slope ($a$), intercept ($b$), total scatter ($\delta_\mathrm{tot}$), total uncertainty ($\delta_\mathrm{tot}$), and intrinsic scatter ($\delta_\mathrm{int}$). The intercept $b$ is obtained starting from the $b'$ value obtained in the fit with mean-subtracted data, as $b=b'+\bar{y}-a\bar{x}$. The total error is derived as the quadratic sum of uncertainties on both axes. The intrinsic scatter considers the scatter along the $y$ variable (i.e., along the excess variance) that is in excess of what can be attributed to the uncertainties on the measurements. For the $\sigma^2$ vs. $M_\mathrm{BH}$ relation, we find a range of 0.65 - 0.73 dex. This is because when we include the 0.40-0.50~dex scatter that comes from single-epoch mass measurements, we obtain a lower intrinsic scatter of $\delta_\mathrm{true}\sim$0.65-0.68 dex (see Section \ref{sec: exvarother}).}
\label{tab:results_MBH_L}
\end{table*}

\section{Dependence on AGN physical properties}
\label{sec: exvarother}

\subsection{Black hole mass}
We first investigated the relationship between the X-ray excess variance and the black hole mass:  
\begin{equation}
    \log(\sigma^2_{100s, 20ks}) + 2.99 = a \cdot (\log(M_{\rm BH}) -7.54)+ b',
\end{equation}
where --2.99 and 7.54 are the mean values for $\log(\sigma^2_{100s, 20ks})$ and $\log(M_{\rm BH})$, respectively. The best-fit parameters are a slope of $a = -0.79 \pm 0.11$ and an intercept of $b' = -0.18 \pm 0.07$ (see Table~\ref{tab:results_MBH_L}). The relation is statistically significant, with $1 - \text{p-value} > 0.99$. The total observed scatter around the best-fit relation is $\delta = 0.77$~dex, calculated using only the detections (i.e., excluding upper limits). Figure~\ref{fig:exvar_Mbh} shows the best-fit relation, along with the median values of $\sigma^2_{\rm NXS}$ in five $M_{\rm BH}$ bins, added to guide the eye. Some of the total scatter is due to the uncertainties on the excess variance and black hole mass values. To estimate how much of the observed scatter can be attributed to measurement uncertainties, we computed the expected uncertainty contribution as $ \delta_{\mathrm{err}} = \sqrt{\delta_{\log \sigma^2}^2 + a^2 \cdot \delta_{\log M_{\rm BH}}^2} = 0.26~\text{dex}$, where $\delta_{\log \sigma^2}$ and $\delta_{\log M_{\rm BH}}$ are the mean uncertainties on the excess variance and black hole mass, respectively. The remaining scatter can be interpreted as an intrinsic component of the relation: $ \delta_{\mathrm{int}} = \sqrt{\delta_{\mathrm{tot}}^2 - \delta_{\mathrm{err}}^2} = 0.73~\text{dex}$. However, the black hole masses used in this work are obtained using the virial equation, with most of the sample having single-epoch (SE) mass measurements, and 32 objects having reverberation mapping (RM) measurements. In both SE and RM cases, black hole masses are derived using the virial relation, $M_{\rm BH} = f \, \frac{R \, \Delta V^2}{G}$, where \( R \) is the radius of the broad-line region (BLR), \( \Delta V \) is the velocity width of the broad emission line, and \( f \) is the virial factor, a dimensionless scaling coefficient that accounts for the geometry, kinematics, and inclination of the BLR. The velocity \( \Delta V \) is usually estimated from the full width at half maximum (FWHM) or the line dispersion of broad permitted lines observed in the rest-frame ultraviolet (e.g., Mg\,\textsc{ii}, C\,\textsc{iv}), optical (e.g., H$\beta$, H$\alpha$), or near-infrared (e.g., Pa$\beta$, Pa$\alpha$) spectral regions \citep[see, e.g.,][]{Greene05, Trakhtenbrot12, Shen13, fricci17}. In RM-based mass measurements, \( R \) is measured directly as the light-travel time delay (lag) between variability in the AGN continuum and the response in the broad emission lines \citep[e.g.,][]{Bentz09}. In SE estimates, \( R \) is inferred from the AGN monochromatic luminosity using the empirical radius--luminosity (\( R\)--\(L \)) relation \citep[e.g.,][]{Kaspi00, Bentz13}, which is calibrated using RM samples.
A major source of systematic uncertainty in both SE and RM mass estimates is the virial factor \( f \), which reflects assumptions about the BLR structure and its orientation relative to the observer. For example, \cite{Ricci_F_2022} reports an anticorrelation of the virial factor \( f \) with the virial mass (likely tracing an underlying anticorrelation with the observed FWHM), as expected if line widths broaden with increasing inclination in a disk-like BLR. The value of \( f \) cannot be measured directly for individual objects and is instead calibrated statistically by matching RM-based AGN black hole masses to the \( M_{\rm BH} \)--\( \sigma_* \) relation observed in quiescent galaxies \citep[e.g.,][]{Onken04, Bennert11, Woo15}. However, the true value of \( f \) may vary substantially across the AGN population due to differences in BLR inclination, turbulence, radiation pressure, and the presence of outflows or inflows \citep{Collin06, Mejia-Restrepo18, Pancoast14}. This introduces an intrinsic uncertainty in mass estimates, typically not captured by the quoted measurement errors.
In the case of SE estimators, additional scatter arises from the use of the \( R \)--\( L \) relation, which carries its own intrinsic dispersion of \(\sim 0.13\)~dex \citep{Bentz13}. Since this relation is empirically derived from RM samples, it assumes that the BLR structure and ionizing continuum shape of the SE objects are similar to those in the RM sample —an assumption that may break down for sources with significantly different accretion properties, luminosities, or BLR conditions. 
These effects contribute to a total population-level uncertainty in virial masses of up to 0.4--0.5~dex \citep{Wandel99, Vestergaard06, Marziani12, Landt13, Shen13, Shen14}, which is significantly larger than formal statistical uncertainties, which typically only account for errors in luminosity and line width.
These systematic uncertainties in the black hole mass estimates inevitably contribute to the observed scatter in the excess variance--mass relation, and should be taken into account when interpreting its parameters. This means that part of the observed $\delta_{\mathrm{int}}$ is still due to this intrinsic mass uncertainty. If we take a value of 0.50~dex for this additional scatter and propagate it through the slope of the best-fit relation, the intrinsic dispersion in the excess variance not attributable to black hole mass uncertainties becomes:
\[
\delta_{\rm true} = \sqrt{\delta_{\mathrm{int}}^2 - a^2 \cdot (0.50)^2} = 0.65~\text{dex}.
\]
If we consider 0.40~dex instead, we obtain $\delta_{\rm true}\sim$0.68 dex. We note that this 0.65-0.68 dex range is only an order of magnitude estimate, as the 0.40-0.50~dex scatter is an average value across AGN populations. The derived best-fit parameters for the $\sigma-M_{\rm BH}$ relation are consistent with the results of \citet{Tortosa23}, as expected given the similarity between the samples, with the present work including fewer upper limits. \\

Our findings are also consistent with recent ensemble-variability results at longer ($\sim$years) timescales. \cite{Georgakakis26} reported, for a sample of $\sim$10000 objects with eROSITA multi-epoch light curves, a clear decrease of the ensemble normalised excess variance with increasing $M_{\rm BH}$ on rest-frame timescales of months to years. Although their methodology and timescale regime differ from the short-timescale excess variance probed in this work, the persistence of the $\sigma^2_{\rm NXS}$–$M_{\rm BH}$ anti-correlation across different regimes reinforces the robustness of the physical connection between black hole mass and X-ray variability amplitude.

In Section \ref{sec:sample}, we described the removal of 10 objects from the sample due to strong evidence of DPE features in their H$\beta$ lines as discussed in \cite{Ward25}, which could bias $M_{\rm BH}$ estimates. To verify whether these objects are genuine outliers, we repeated the fit with the full sample, including the DPE-flagged objects. The scatter of the subsample was found to be $\delta = 0.89$~dex. This corroborates their removal, as likely their masses are unreliable.

We also tested the relation between the excess variance and the $M_{BH}$ restricting it to the RM sample, which is shown on the right panel of Fig. \ref{fig:exvar_Mbh}. The slope we obtain is steeper than for the whole sample, but still consistent within uncertainties. Similar results were found by \cite{Tortosa23} (see their Section 3.2). Although RM masses are more reliable, the sample size is less than one third when compared to the whole sample, which contributes to the overall higher uncertainties in the relation parameters estimates, although uncertainties on $M_{BH}$ are smaller. Future, more homogeneous RM datasets will allow a more robust comparison. For the purposes of this work, the RM-only and full-sample relations are statistically consistent, and the inclusion of the RM‑only result does not alter our conclusions.

\subsection{Luminosity}
\begin{figure*}[htbp]
    \centering
    \includegraphics[width=0.46\textwidth]{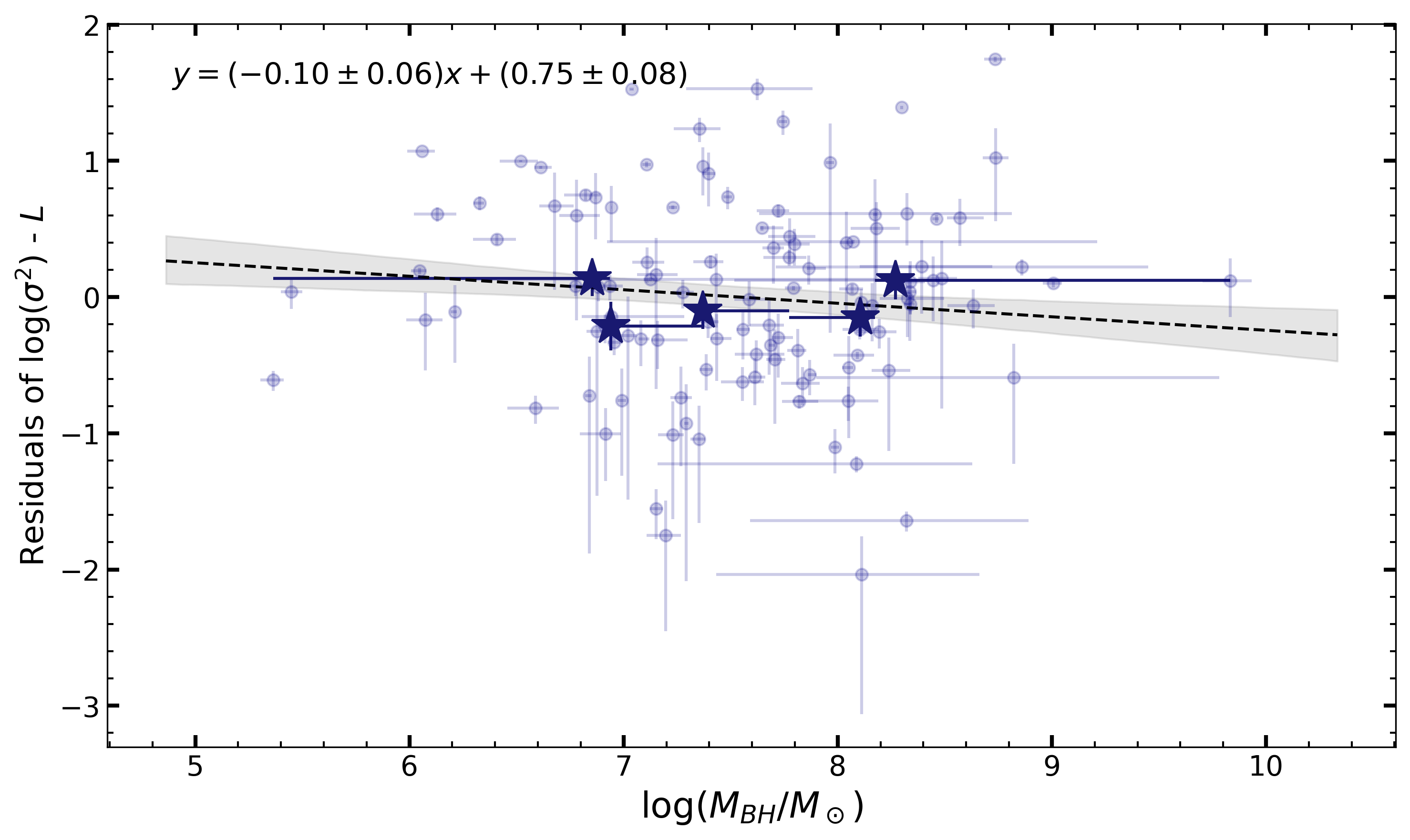}
    \hfill
    \includegraphics[width=0.46\textwidth]{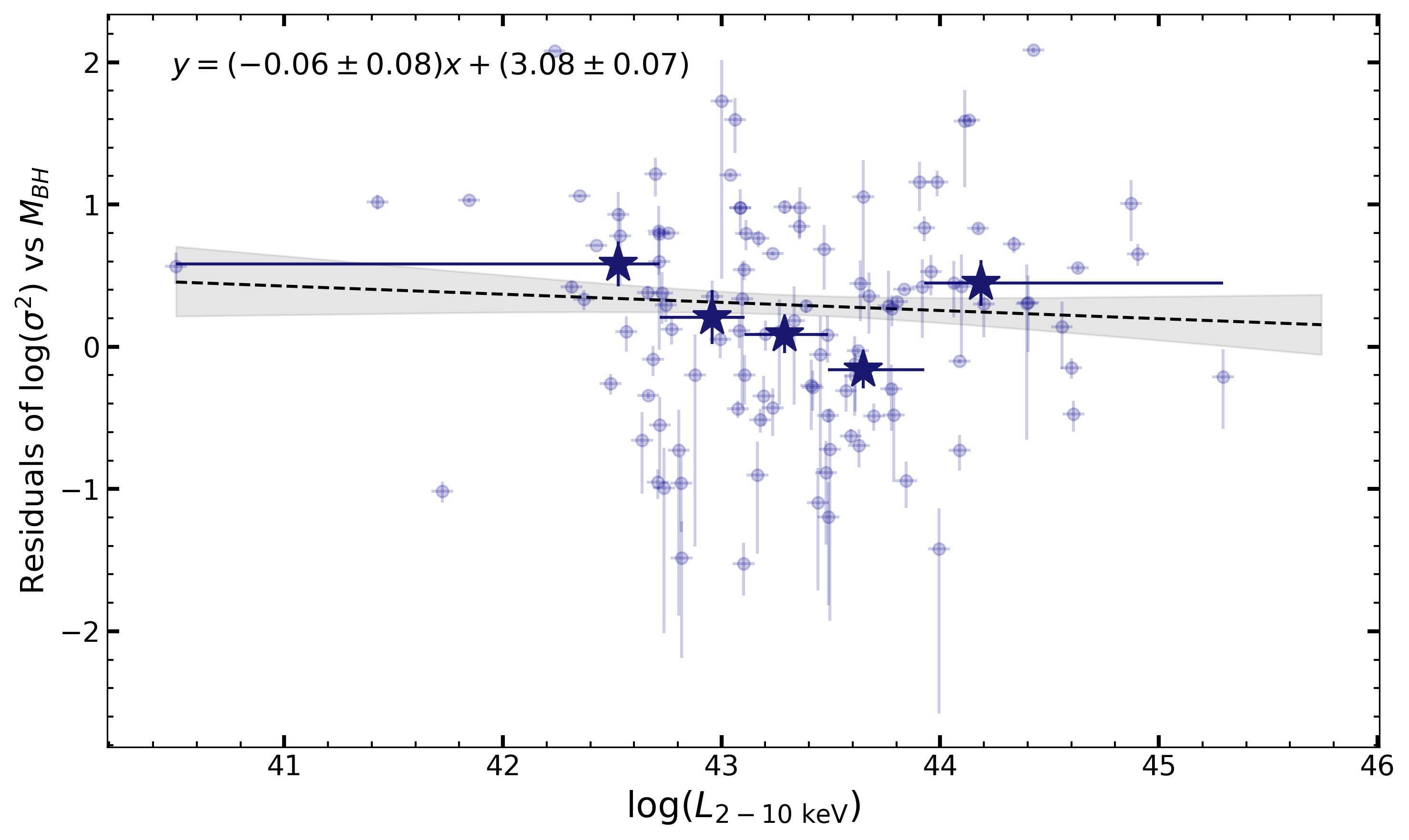}
    \includegraphics[width=0.46\textwidth]{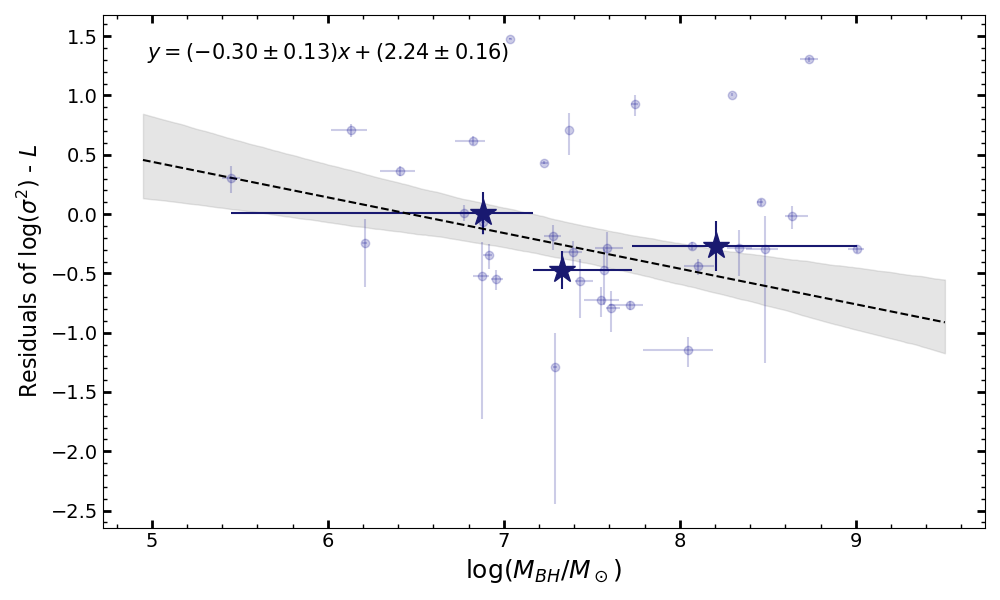}
    \hfill
    \includegraphics[width=0.46\textwidth]{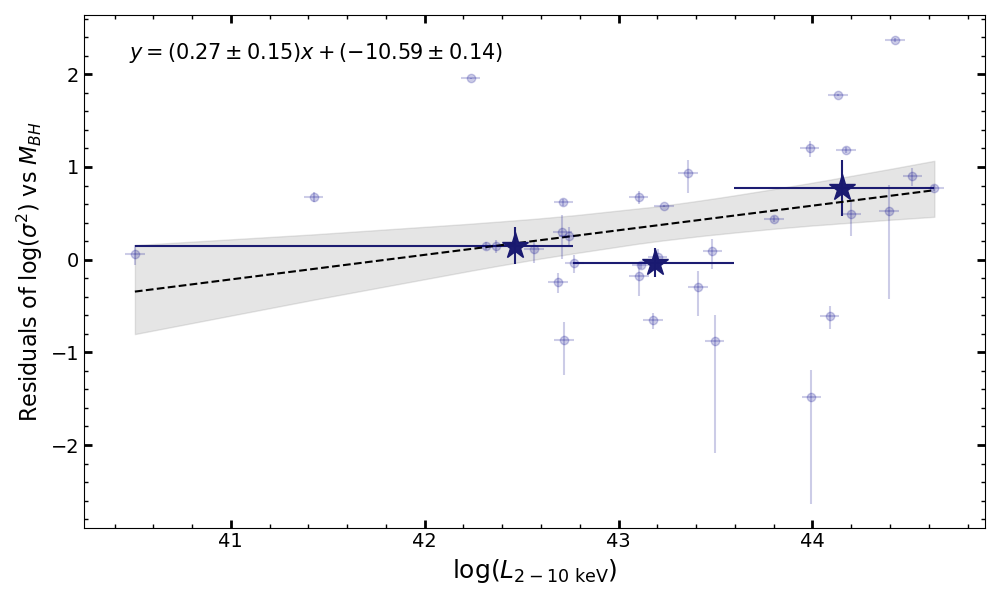}
    \caption{
    Residual analysis to assess whether black hole mass or luminosity more fundamentally drives X-ray excess variance. \textit{Upper left panel}: residuals from the best-fit of the $\log(\sigma^2)$–$\log(L_{2-10\,\mathrm{keV}})$ relation plotted against $\log(M_{\rm BH})$. A weak trend is observed ($\alpha = -0.10 \pm 0.06$), but the obtained p-value is 0.88, against the significance of the trend. \textit{Upper right panel}: residuals from the $\log(\sigma^2)$–$\log(M_{\rm BH})$ relation plotted against $\log(L_{2-10\,\mathrm{keV}})$, showing no significant trend ($\alpha = -0.06 \pm 0.08$); the p-value is 0.75. These findings suggest that for this dataset, neither $M_{\rm BH}$ nor $L$ can be considered the 'main driver' of the variability against the other. \textit{Lower left panel}: the same as in the Upper left panel, but restricted to the RM-only sample. A 2.3 $\sigma$ significant negative dependence on $M_{BH}$ is observed.  \textit{Lower right panel}: the same as in the Upper right panel, but restricted to the RM-only sample. A 2 $\sigma$ positive dependence on luminosity is observed. This marginally suggests that $M_{BH}$ is more predictive than the luminosity.}    
    \label{fig:residual}
\end{figure*}

We explored the relationship between the X-ray excess variance and intrinsic luminosity, testing the 2–10 keV X-ray luminosity and the monochromatic optical luminosities at 5100~\AA{} and 6200~\AA{}. The 2-10 keV X-ray luminosities are from the \cite{Ricci_C_17} catalogue, where they are derived from X-ray spectral fitting. The results are summarised in Table \ref{tab:results_MBH_L} and illustrated in Figure \ref{fig:exvar_L}. The best-fit parameters for the 2–10~keV luminosity are consistent with the findings of \citet{Tortosa23}, as expected. Interestingly, the slopes derived for the optical luminosities are flatter compared to those for the X-ray luminosity, although being consistent within 1-$\sigma$. This is indeed expected given the well-known nonlinear relation between the X-ray and the UV luminosities in AGN \citep[e.g.,][]{Tananbaum79, Vignali03, Lusso10}. The total dispersion values range from $\delta = 0.73$~dex to $\delta = 0.78$~dex, with intrinsic scatter values similar to those obtained for the $M_{\rm BH}$ relation, approximately $\delta_{\mathrm{int}} \sim 0.70$~dex. \\
We compare our results with those of \citet{LaFranca14}, who investigated the relationship between the excess variance and luminosity in both the X-ray (2–10 keV) and optical (5100~\AA) bands. Their work employed a sample of 40 objects from \citet{Ponti12}, with no upper limits, and with distribution in redshift (0$<z<0.3$) and 2-10 keV luminosities ($\log(L_{2-10 keV}/erg s^{-1})$ $\sim$ 40-45) analogous to those of the BASS sample. They fitted the relation $\log(L) = a' \cdot \log(\sigma^2) + b'$, obtaining slopes of $a'_{2-10\,\mathrm{keV}} = -1.67$ and $a'_{5100\,\mathrm{\AA}} = -1.98$. These results are consistent with ours, as the slope in our formulation ($a$) is the reciprocal of their slope ($a'$). Specifically, $1/(-1.67) \sim -0.60$ and $1/(-1.98) \sim -0.51$. So, the increased statistics of the BASS samples still provide results consistent with the previous ones in the literature. Regarding the dispersion of the relation, we will focus on that in Section \ref{sec: distances}.

\subsection{What is the main driver of X-ray variability?}
Identifying the primary driver of AGN X-ray variability is key to understanding the underlying physical mechanisms. Both black hole mass and luminosity are known to correlate with the X-ray excess variance, but determining which of the two plays a more fundamental role remains an open question. One possible approach is to compare the intrinsic scatter of the two relations: a lower scatter can indicate a tighter, possibly more physically rooted correlation. This is indeed how previous works have suggested that the main driver of X-ray variability is the $M_{\rm BH}$, with the relation with the luminosity being a consequence of it \citep[e.g.,][]{Ponti12}. In our sample, we find that the total observed scatter in the $\sigma^2$–$M_{\rm BH}$ relation ($\delta = 0.77$~dex) is consistent with that in the $\sigma^2$–$L$ relation (ranging from $\delta = 0.73$ to $0.76$~dex). However, this comparison is complicated by the nature of the black hole mass estimates, which are primarily based on single-epoch virial methods. As discussed in the previous Section, when we account for this additional scatter, we find that the intrinsic scatter in the $\sigma^2$–$M_{\rm BH}$ relation drops to $\sim$0.65–0.68~dex, thus becoming comparable to or even smaller than that of the luminosity-based relation. This gives hint that, despite the larger observed dispersion, black hole mass may still be more fundamentally linked to variability, and the higher scatter is, at least in part, an artifact of mass measurement uncertainties. 
Additional information can come from the value of the slopes, as the steepness of the relation can indicate how rapidly the variability amplitude responds to changes in each predictor. In our sample, the best-fit slopes in the $\log\sigma^2$–$\log M_{\rm BH}$ and $\log\sigma^2$–$\log L_{2-10\,\mathrm{keV}}$ relations are comparable within $1\sigma$ (Table~1), suggesting similar response strengths. In the RM-only subset, the $\log\sigma^2$–$\log M_{\rm BH}$ slope is steeper, yet still consistent within uncertainties. In the RM-only subset, the $\log\sigma^2$–$\log M_{\rm BH}$ slope is steeper, yet still consistent, within uncertainties, with the whole sample.
To further disentangle the contributions of $M_{\rm BH}$ and $L$, we performed a residual analysis. This method tests whether one parameter adds predictive power beyond what is already explained by the other. Specifically, if the residuals of the $\log(\sigma^2)$–$\log(L_{2-10\,\mathrm{keV}})$ relation correlate with $\log(M_{\rm BH})$, it would imply that $M_{\rm BH}$ captures additional variability-related information not encoded in luminosity alone.\\
We first computed the residuals from the $\log(\sigma^2)$–$\log(L_{2-10\,\mathrm{keV}})$ best-fit relation and regressed them against $\log(M_{\rm BH})$, obtaining a p-value of 0.88 and slope of $\alpha = -0.10 \pm 0.06$. The trend is consistent with zero at a 2$\sigma$ level, while the high p-value tells us that there is no significant trend. The residuals from the $\log(\sigma^2)$–$\log(M_{\rm BH})$ relation show no dependence on $\log(L_{2-10\,\mathrm{keV}})$, with a p-value 0.75 and a slope of $\alpha = -0.06 \pm 0.08$. Again, the high p-value obtained here can be considered as proof against the presence of a significant trend in the residuals. These results are shown in the upper panels of Figure~\ref{fig:residual}. 
We also performed the same analysis but restricted to the RM-only sample, as shown in the lower panels of Figure~\ref{fig:residual}. The analysis shows weak trends ($\sim2\sigma$) in opposite directions depending on which variable is held fixed. The trend with $M_{\rm BH}$ residuals is slightly more significant, at a 2.3$\sigma$ level. This would suggest the $M_{\rm BH}$ as the more relevant driver, but the result is only marginal.

Taken together, the results on the scatter and the slight trend in $\log(\sigma^2)$–$\log(L_{2-10\,\mathrm{keV}})$ residuals may suggest that $M_{\rm BH}$ is more closely related to the physical mechanism driving X-ray variability than luminosity. However, the statistical evidence remains very tentative. Ultimately, these findings highlight the importance of improving black hole mass measurements in large AGN samples. Future studies based on reverberation mapping or improved calibrations of virial estimators, especially those accounting for systematic dependencies of the virial factor, will be critical for resolving the relative contributions of mass and luminosity in determining AGN variability \citep[e.g.,][]{Villafana23, Villafana24}.

\subsection{Eddington ratio} 
\label{subsec: eddratio}
\begin{figure*}[htbp]
    \centering
    \includegraphics[width=0.47\textwidth]{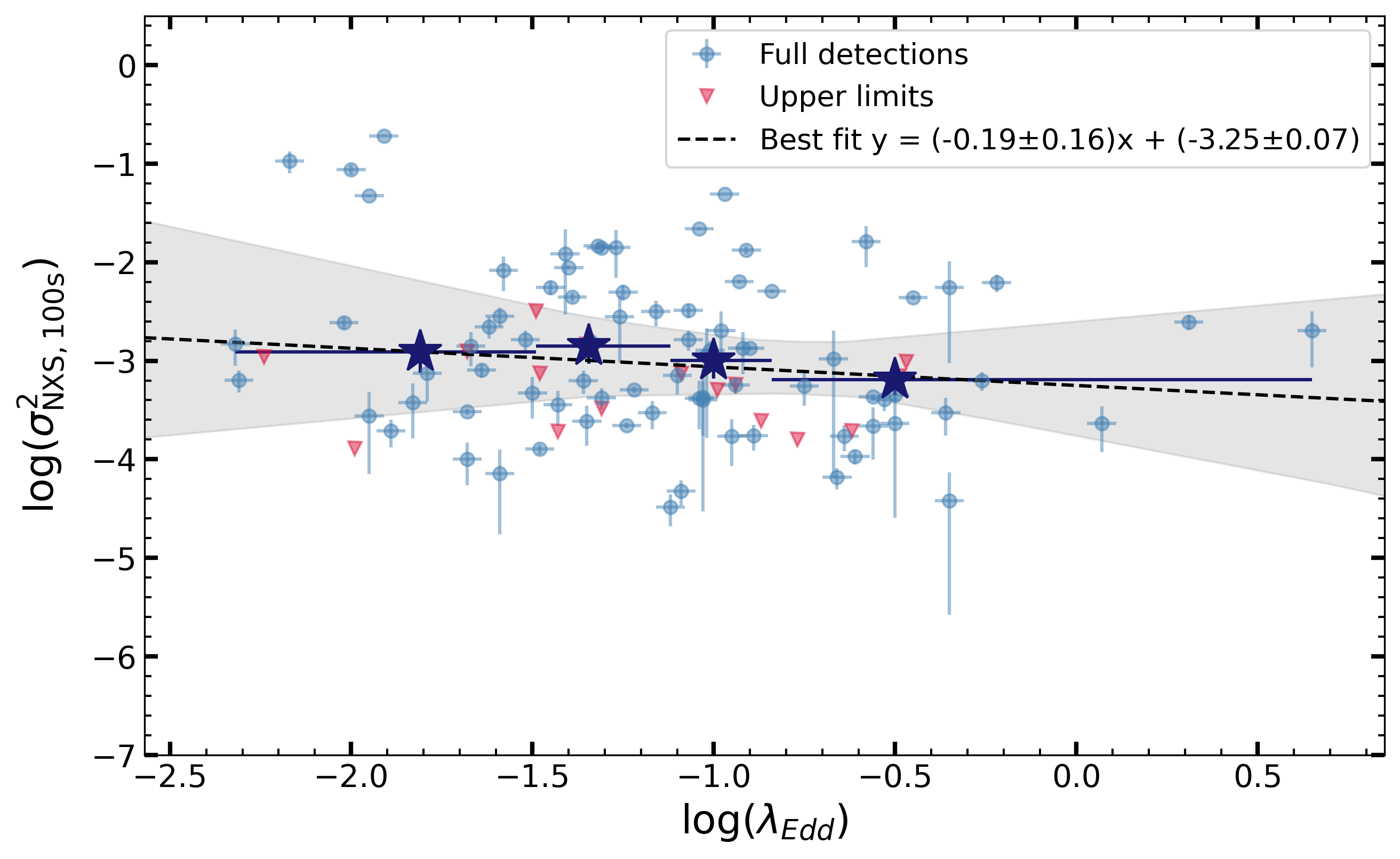}
    \hfill
    \includegraphics[width=0.47\textwidth]{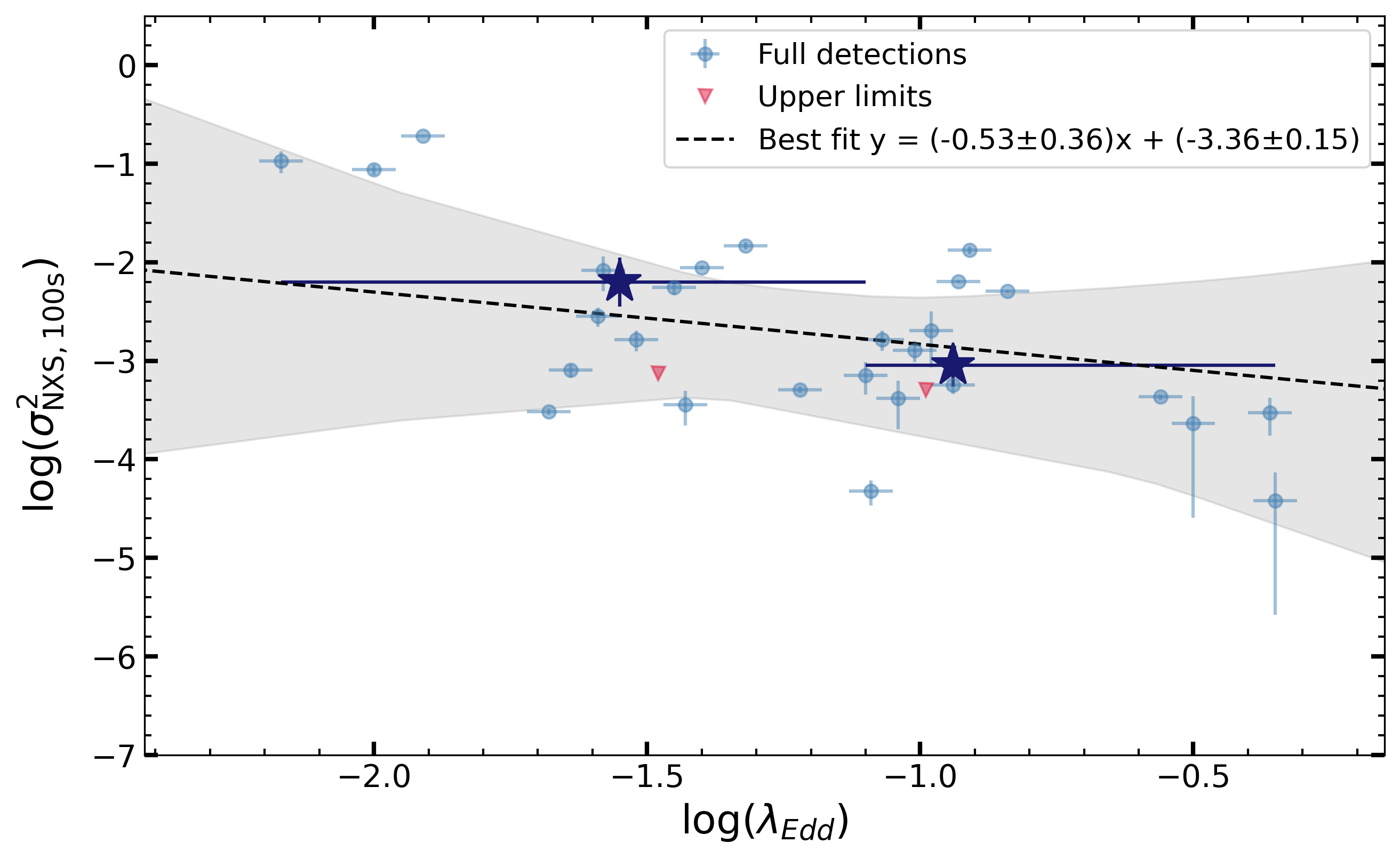}
    \caption{Relation between the X-ray excess variance $\sigma_{\rm NXS}$ and the $\lambda_\mathrm{Edd}$. Blue stars represent averaged bins in Eddington ratio and are shown to guide the eye.
    Left panel: fit performed for the whole sample, where 102 objects out of 134 have black hole masses estimated using the virial equation, and 32 have Reverberation Mapping estimates. Despite the reduced fraction of upper limits and increased precision of the $\lambda_\mathrm{Edd}$ estimates compared with other studies, we still see no significant relation between X-ray variability and the $\lambda_\mathrm{Edd}$. Right panel: same analysis for the subsample of objects with black hole masses derived from Reverberation Mapping (N=32). We see an anti-correlation, but it is slightly significant.}
    \label{fig:exvar_eddratio}
\end{figure*}

The relationship between the X-ray excess variance and the $\lambda_\mathrm{Edd}$ is of significant interest, as it can provide insight into the variability properties of AGN. According to \citet{McHardy06}, the break timescale in the PSD increases with $M_\mathrm{BH}$ and decreases with increasing $\lambda_\mathrm{Edd}$. If AGN follow a universal PSD characterised by a single break frequency dependent on $M_\mathrm{BH}$, combined with a constant normalisation on long timescales, one would expect the strength of the relations $\sigma^2$ vs. $M_\mathrm{BH}$ and $\sigma^2$ vs. $\lambda_\mathrm{Edd}$ to be comparable \citep{Sartori18}. This would predict higher short-timescale variability for AGN with lower $M_\mathrm{BH}$ and higher $\lambda_\mathrm{Edd}$. However, many studies have not identified a significant correlation between variability and the $\lambda_\mathrm{Edd}$. For example, \citet{O'Neill05}, \citet{Gierlinski08}, \citet{Zhou10}, \citet{Ponti12}, and \citet{Lanzuisi14} reported no robust evidence of such a relationship. Similarly, for the BASS sample, \citet{Tortosa23} found no statistically-significant correlation. 
In this work, we aim to re-examine this relationship under updated conditions that differ from \citet{Tortosa23}. First, our sample includes fewer upper limits, which should increase the sensitivity to any underlying relation. Second, we removed objects exhibiting signs of DPE in the H$\beta$ lines, as these objects likely have biased $M_{\rm BH}$ estimates, which would propagate to their $\lambda_{\rm Edd}$ estimates. Finally, we now use $\lambda_\mathrm{Edd}$ values derived from bolometric luminosities obtained through SED fitting \citep{Gupta24}. 
We tested the relation
\begin{equation}
    \log(\sigma^2_{100s, 2-10\,\mathrm{keV}}) + 2.99 = a \cdot (\log(\lambda_\mathrm{Edd}) + 1.27) + b',
\end{equation}
where -2.99 and -1.27 are the mean values of $\log(\sigma^2_{100s, 2-10\,\mathrm{keV}})$ and $\log(\lambda_\mathrm{Edd})$, respectively,
and performed the fit as discussed in Sect. \ref{sec: fitting}. The results are shown in the left panel of Figure \ref{fig:exvar_eddratio}. For visualisation purposes, we added blue stars representing the median $\lambda_\mathrm{Edd}$ values in four bins, spaced to ensure that each bin contains at least 20 objects. The p-value of the correlation is 0.16, and the best-fit slope is $a = -0.14 \pm 0.15$, consistent with zero, indicating that the relationship is not significant. Therefore, even with this improved sample, including fewer upper limits and updated $\lambda_\mathrm{Edd}$ values, we still do not find evidence for a significant correlation between excess variance and the $\lambda_\mathrm{Edd}$.\\
It is still possible that the relationship is weak but exists, hidden by large uncertainties, again mainly due to black hole mass estimates. To mitigate this, we restricted our analysis to a subset of 32 objects with $M_\mathrm{BH}$ obtained via RM, which are considered more accurate than single-epoch mass measurements. For this subset, we fit the same relation as above, and the results are shown in the right panel of Figure \ref{fig:exvar_eddratio}. The slope in this case is $a = -0.53 \pm 0.36$, with a p-value of 0.003, indicating a statistically significant, though weak, correlation.
Despite the statistical significance, this result should be interpreted with caution. The subsample of RM objects is small and may not be fully representative of the broader AGN population. Objects targeted for RM campaigns are often less massive and exhibit greater variability, as this increases the likelihood of detecting time lags and enabling $M_\mathrm{BH}$ measurements during the monitoring campaign. For our sample, the difference in mean $M_\mathrm{BH}$ is modest, with $\langle \log(M_\mathrm{BH}/M_\odot) \rangle = 7.6 \pm 0.8$ for non-RM objects and $\langle \log(M_\mathrm{BH}/M_\odot) \rangle = 7.4 \pm 0.8$ for RM objects. Nevertheless, this difference could still contribute to the observed discrepancy. Further studies with larger, unbiased samples with robust $M_\mathrm{BH}$ and $\lambda_{\rm Edd}$ are required to clarify the nature of the relationship between excess variance and $\lambda_\mathrm{Edd}$.
As an additional test, we also explored the relation between the excess variance and the “X‑ray Eddington ratio”, $L_X/L_{Edd}$, as shown in Appendix \ref{app:Xrayedd}. This quantity can in principle offer further insight into how variability depends on the relative strength of the X‑ray emission. However, the correlation is only marginally significant, and strong degeneracies with black hole mass and luminosity are likely at play. For these reasons, and to avoid over‑interpreting the result, we present the full discussion in the Appendix.

\section{Cosmological implementation}
\label{sec: distances}
Motivated by the established link between X-ray excess variance and luminosity, we test whether AGN variability can serve as a distance indicator. Because $\log(\sigma^2_{100s, 20ks})$ can, in principle, provide a cosmology-independent estimate of intrinsic luminosity, this approach could extend standard-candle work beyond the redshift reach of Type Ia supernovae. In this Section, we evaluate this idea using the BASS sample, inverting the perspective adopted above: rather than treating $\log(\sigma^2_{100s, 20ks})$ as the dependent variable, we use it as the predictor of intrinsic luminosity.
Therefore, we now treat luminosity as the dependent variable in the relation:
\begin{equation}
\log (L) - \overline{\log (L)}= \alpha (\log(\sigma^2_{100s, 20ks}) -\overline{\log(\sigma^2_{100s, 20ks}})) + \beta.
\label{eq_for_D}
\end{equation}
We attempted to fit the different luminosities in Eq. \ref{eq_for_D}: the 2-10 keV luminosity, the monochromatic 5100~\AA ~luminosity and the 4400~\AA~  luminosity. We found that using X-ray luminosities gave the best proxy for the excess variance, with the 2-10 keV luminosity showing the lowest scatter (0.63 dex), compared to 0.70 dex and 0.72 dex for the 4400~\AA~and 5100~\AA~luminosities. This difference can be explained by the fact that the X-ray and optical observations of the sample were not simultaneous. As AGN emission can vary on short timescales, it could be that the X-ray corona state was different when optical observations were performed. The 2-10 keV luminosity is more directly tied to the X-ray excess variance as it is measured from the same observation.
Note that the dispersion in this inverse relation is not directly comparable to the scatter discussed earlier and presented in Table \ref{tab:results_MBH_L}. For cosmological purposes, we will focus on the relation with X-ray luminosity, which has the lowest scatter. The found total scatter of 0.63 dex in the $\log(L_{2-10 \, \rm{keV}}) - \log(\sigma^2_{100s, 20ks})$ relation, which is substantially lower than the 1.33 dex reported by \citet{LaFranca14}. We note that the sample in \citet{LaFranca14} has analogous distribution in redshift, $M_{BH}$ and luminosity as the BASS sample of this work. The reduced scatter is likely due to both the larger sample size (134 vs. 40 sources) and improved completeness. This reduction is crucial, as the dispersion directly limits the precision of distance estimates derived from this relation.

Given an estimate of the intrinsic luminosity from Eq.~\ref{eq_for_D}, we can compute the luminosity distance using:
\begin{equation}
D_L = \sqrt{\frac{L_{2-10 \, \mathrm{keV}}}{4 \pi F_{2-10 \, \mathrm{keV}}}},
\end{equation}
where $F_{2-10 \, \mathrm{keV}}$ is the observed flux. Propagating the uncertainty on $L$ through this relation gives the uncertainty on the luminosity distance: $\delta_{\log D_L} = \frac{1}{2} \delta_{\log L_{2-10 \, \mathrm{keV}}} = 0.315 \, \mathrm{dex}$.

In linear terms, this corresponds to a fractional uncertainty:
\begin{equation}
\frac{\Delta D_L}{D_L} = \frac{\delta_{\log D_L}}{\log (e)} = \frac{\delta_{\log D_L}}{0.434} \approx 0.73 \, \text{or } 73\%.
\end{equation}
This is a substantial improvement over previous results, but still larger than the uncertainties achievable with other cosmological probes. For reference, Type Ia supernovae typically yield 5–10\% uncertainties, while quasar-based $L_{\rm X}$–$L_{\rm UV}$ relations can achieve 12–20\% \citep{Lusso20, Signorini23b}.

\citet{LaFranca14} proposed a strategy to reduce the scatter in variability-based luminosity estimates by including the FWHM of broad emission lines. The idea is based on the virial relation:
$ M_{\rm BH} = f \, \frac{R \, \Delta V^2}{G}$, and on the assumption that black hole mass is a fundamental driver of variability. If true, including line width could help recover this deeper physical link and reduce dispersion. In their work, \citet{LaFranca14} found that the relation
\begin{equation}
\log(L_{2-10 \, \mathrm{keV}}) + 4 \log(\mathrm{FWHM}) = \alpha  \log(\sigma_{NXS}^2) + \gamma
\label{eq:with_fw}
\end{equation}
yielded a smaller dispersion than Eq.~\ref{eq_for_D} alone. In our sample, however, the variability–mass and variability–luminosity relations show comparable observed scatters, as discussed in Section~\ref{sec: exvarother}. This suggests that the variability is not more fundamentally linked to mass than to luminosity, at least within our sample. Consequently, we do not expect significant improvements by including FWHM in the luminosity prediction.

\begin{figure}[htbp]
    \centering
    \includegraphics[width=0.46\textwidth]{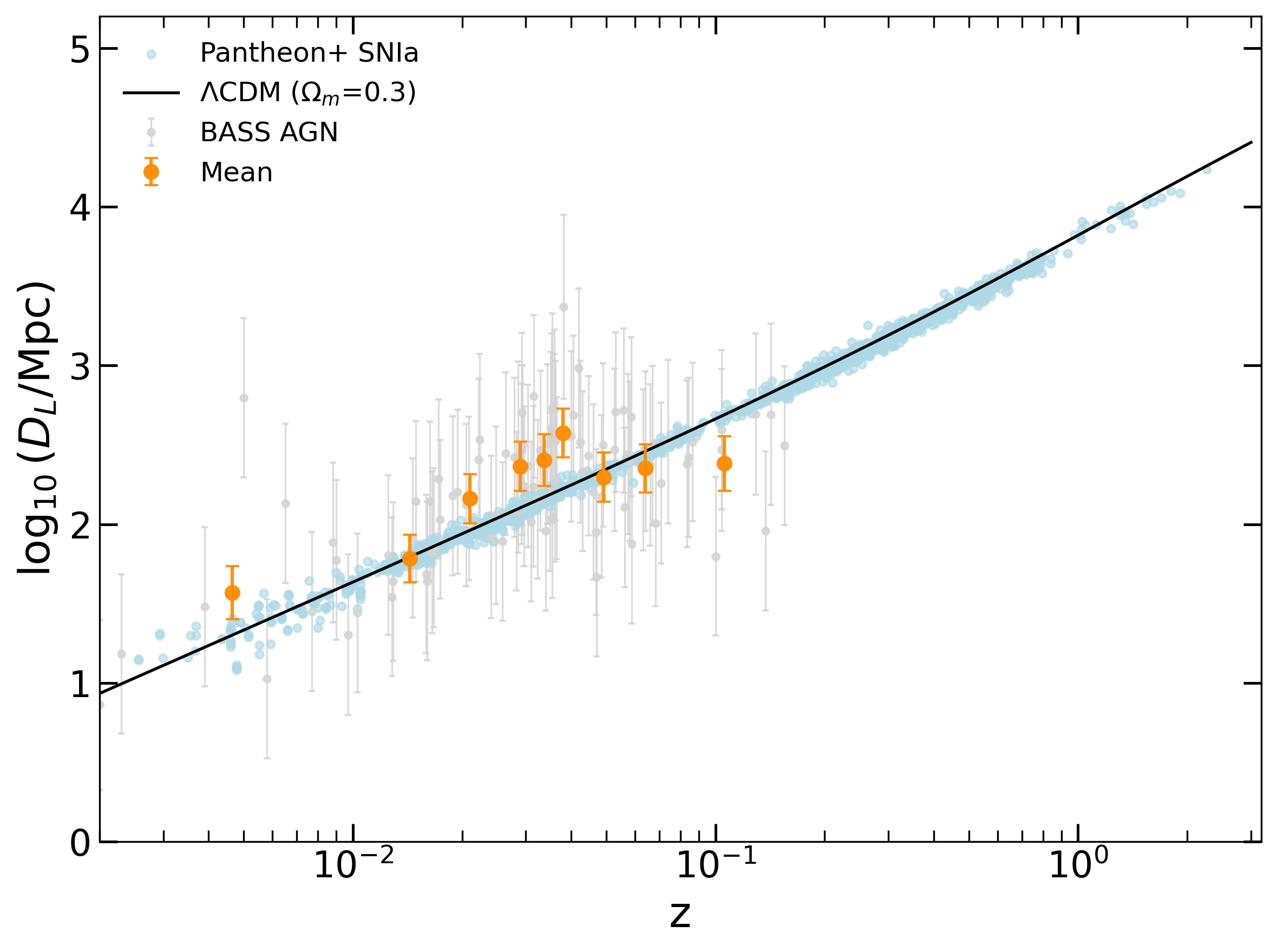}
    \caption{
   Proof-of-concept Hubble Diagram for the BASS sample (gray points) using the relation between X-ray excess variance and the 2-10 keV luminosity. The dark orange points show the mean values obtained by binning the sample into nine equal-sized bins. Cyan points represent the Pantheon+ Supernovae Ia sample \citep{Scolnic18}, while the black line represents the prediction from a standard flat $\Lambda$CDM model. The redshift range and limited statistics of the BASS sample currently undermine the cosmological implementation of the method, but future probes might open new possibilities (see Section \ref{sec: future}).
    }
    \label{fig:HD_BASS}
\end{figure}

For completeness, and to enable direct comparison with \citet{LaFranca14}, we repeated their analysis on our larger sample. The results are presented in Appendix~\ref{app: fwhm}. We find no reduction in scatter when using the line widths of H$\alpha$ or H$\beta$. A moderate decrease is observed when using Pa$\alpha$ and Pa$\beta$, but the sample sizes remain limited (N = 40 for Pa$\alpha$, N = 51 for Pa$\beta$). The inclusion of FWHM may be effective only when the initial dispersion is large; in our case, the already reduced scatter from sample improvements may leave little room for further improvement. As a proof of concept, we show the Hubble Diagram of the BASS sample in Figure \ref{fig:HD_BASS}. Objects are shown in light gray; dark orange points are the mean values of the luminosity distances when dividing the sample into nine bins of equal size. For a comparison, the distance values for SNIa from the Pantheon+ sample \citep{Scolnic22} are shown in cyan, and the shape of the Hubble Diagram predicted by a flat $\Lambda$CDM model is shown in solid black. Clearly, the limited redshift range and statistics of the BASS sample make it a worse probe than SNIa for measuring cosmological distances. However, AGN can be observed at much higher redshifts than SNIa, and future X-ray observatories could be of great importance for measuring distance using X-ray variability, as discussed in the next section. 

\section{Future observatories}
\label{sec: future}

\begin{figure*}[htbp]
    \centering
    \includegraphics[width=0.49\textwidth]{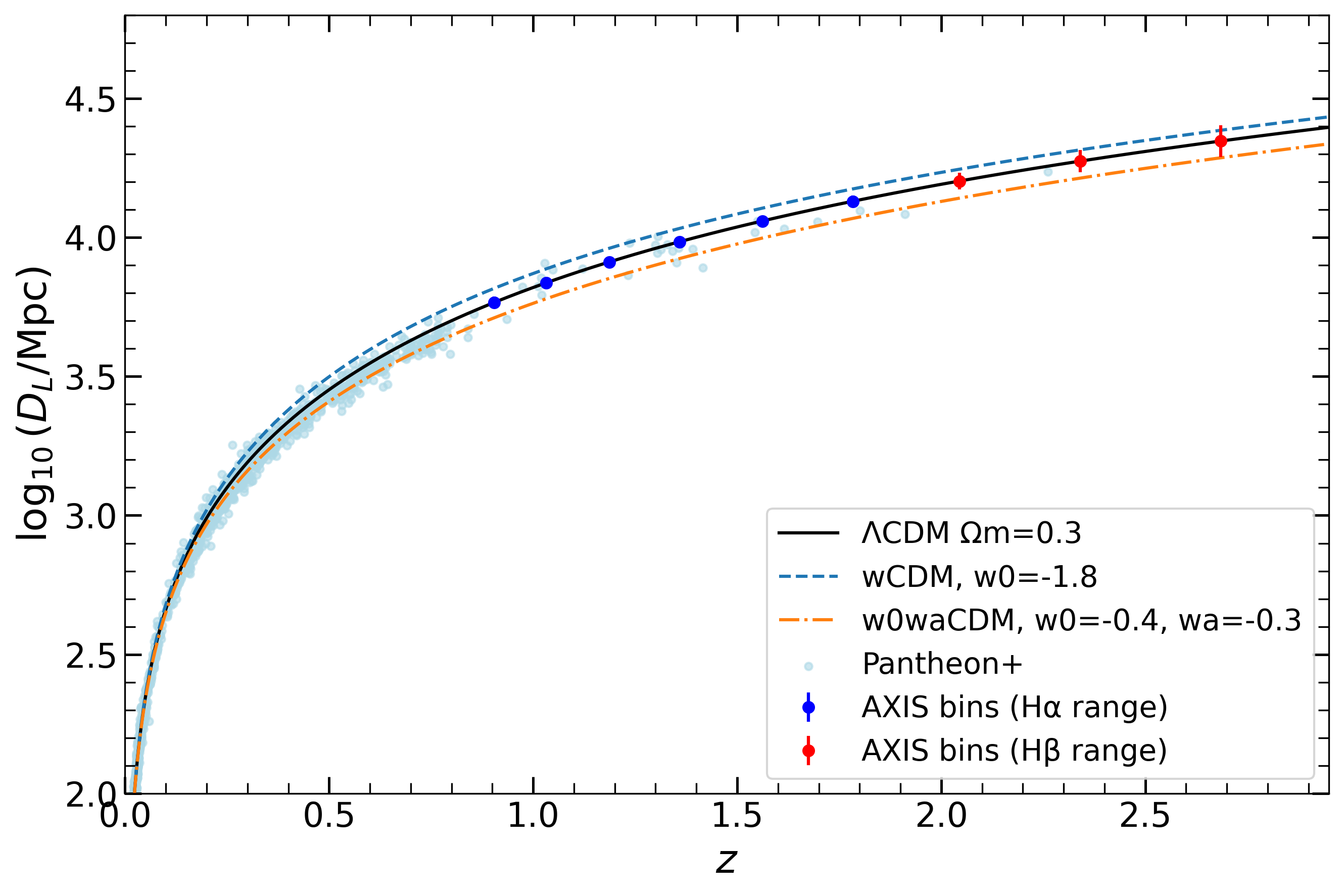}
    \hfill
    \includegraphics[width=0.49\textwidth]{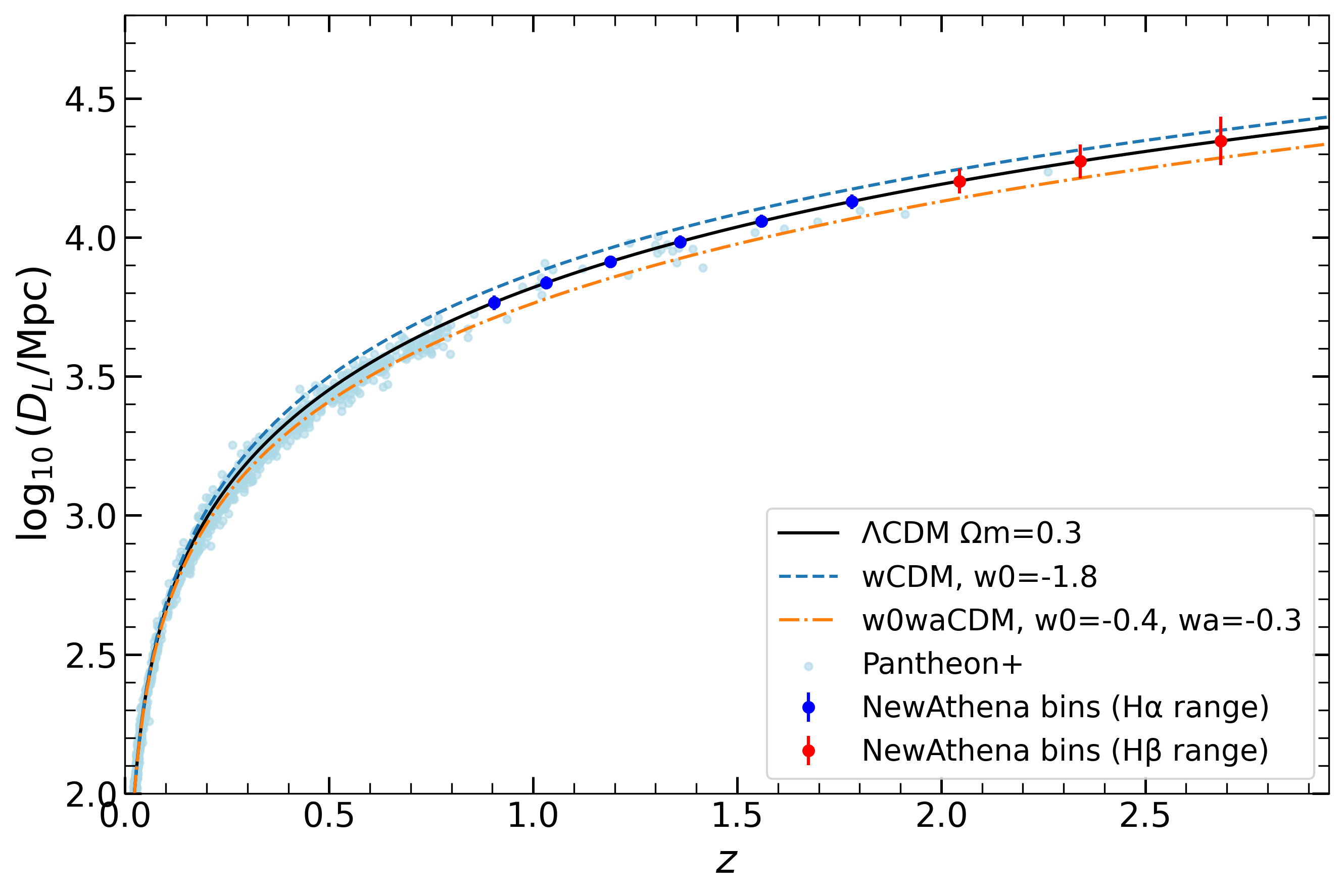}
    \caption{
    Left: simulated Hubble Diagram obtained using the relation between X-ray excess variance and 2--10~keV luminosity to derive the luminosity distance $D_L$, for sources expected to be observed by \textit{AXIS} (to measure variability) and \textit{Euclid} (to provide spectroscopic redshifts). 
    Right: same simulation for \textit{NewAthena}, considering its planned deep survey coverage. 
    The light cyan points represent the luminosity distances obtained from SNe~Ia in the Pantheon+ sample. 
    The black solid, dashed, and dot--dashed curves correspond to different cosmological models, which overlap at low redshift but diverge at $z\gtrsim2$, where AGN become valuable standardisable candles. 
    }
    \label{fig:HD}
\end{figure*}

The current results on the relation between X-ray excess variance and luminosity still exhibit significant dispersion, limiting their immediate use as a cosmological distance indicator (and therefore as a means to distinguish between cosmological models). However, future X-ray observatories will dramatically increase the number of AGN with precise excess variance measurements, enabling tighter calibrations of this relation and its cosmological implementation. 

\textit{AXIS}\footnote{\url{https://axis.umd.edu/}} \citep{Reynolds23}, a NASA concept currently in Phase~A, will provide sub-arcsecond imaging with a large effective area and field of view, ideal for serendipitous AGN variability studies. 
\textit{NewAthena} \citep{Cruise25_newathena}, ESA next-generation X-ray observatory expected to be adopted by 2027, will host the Wide Field Imager (WFI), featuring a 40’$\times$40’ field of view and an energy resolution of $<$170 eV at 7 keV, and will offer a resolving power comparable to \xmm but with a considerably larger collecting area.\footnote{\url{https://www.mpe.mpg.de/ATHENA-WFI/}} Both missions will therefore open a new regime for precision AGN variability measurements.

To estimate how many AGN could be used as cosmological probes with these missions, we start from the 100~deg$^2$ mock AGN population of \citet{Gilli07}, which represents a realistic census of sources over a wide redshift and luminosity range and also accounts for the expected fraction of AGN that would be obscured.   
We then apply our selection criteria to identify objects that could be observed either by \textit{AXIS} or \textit{NewAthena}, and that would yield reliable X-ray excess variance measurements. 
For \textit{AXIS}, we consider the entire 100~deg$^2$ mock field as a realistic proxy for all serendipitous AGN that will accumulate over the mission lifetime. 
For \textit{NewAthena}, the current observational strategy includes 70~pointings of 200~ks and 30~pointings of 300~ks each (private communication, \textit{NewAthena} team). It has not been decided yet exactly which fields will be covered by these surveys, but the total would amount to a sky area of 44 deg$^2$. We therefore consider the numbers from the 100~deg$^2$ catalog multiplied by 0.44. 

As shown in our BASS sample, at least 40~counts in the rest-frame 2--10~keV band over a 20~ks observation are required to obtain a robust measurement of the excess variance. 
Because of cosmological time dilation, a rest-frame 20~ks observation corresponds to $20~\mathrm{ks}\times(1+z)$ in the observed frame, and we therefore require each object to have a total exposure of $20~\mathrm{ks}\,(1+z)$ and at least $40\,(1+z)$ counts. 
For \textit{NewAthena}, the long planned exposures easily satisfy this condition even at the highest redshifts.
For \textit{AXIS}, although some serendipitous fields may have shorter exposures, the mission’s expected lifetime and cumulative coverage make it plausible that more than 100~deg$^2$ will eventually be observed to sufficient depth. Given \textit{AXIS} and \textit{NewAthena} energy ranges, that are of 0.3–10 keV and 0.2–15 keV respectively, the rest-frame 2--10~keV band will be directly observed for all objects in the redshift range 0$<z<5.7$ for \textit{AXIS} and 0$<z<9$ for \textit{NewAthena}.

We then use the latest response matrices for both telescopes through \textsc{WebPIMMS}\footnote{\url{https://heasarc.gsfc.nasa.gov/cgi-bin/Tools/w3pimms/w3pimms.pl}} to convert the 2--10~keV flux into count rate for an unobscured AGN with a photon index $\Gamma=1.8$. We apply these conversions to the 100~deg$^2$ mock catalog, selecting only sources that are unobscured, which means with a column density of $\log(N_{\mathrm{H}}/{\rm cm}^{-2})<21$, and bright enough to give a reliable X-ray excess variance measurement. We performed simulations over a photon index range of $\sim1.7-2.0$, which is reasonable for unobscured AGN, and obtained results similar to those obtained. For simplicity, we discuss the results for the $\Gamma=1.8$ prescription. 

We also need these objects to have a reliable spectroscopic redshift measurement. This step is crucial for the cosmological application, as reliable distances can only be derived when both variability and redshift are measured. For the sake of this exercise, we consider how many will have a detectable emission line with \textit{Euclid}, the ESA space mission designed to map the geometry of the dark Universe using weak lensing and galaxy clustering \citep{Laureijs11}. 
\textit{Euclid} is equipped with two instruments, VIS \citep{ECCropper25} and the Near-Infrared Spectrometer and Photometer \citep[NISP-S and NISP-P,][]{ECJahnke25}. Over its 6-year survey duration, it will map about one-third of the extragalactic sky in the \textit{Euclid} Wide Survey (EWS) \citep{ECScaramella22}. During the EWS, slitless spectroscopic data will be acquired by NISP-S using the red grism RG$_{\rm E}$, covering 1206–1892~nm.
This wavelength range allows the detection of H$\alpha$ at redshifts between $\sim$0.84  and $\sim$1.9, and H$\beta$ between $\sim$1.9 and $\sim$2.9. From the \cite{Lusso24} catalog, the flux limit for detecting H$\alpha$ with \textit{Euclid} is approximately $5 \times 10^{-16}$ erg cm$^{-2}$ s$^{-1}$. Using the correlation from \citet{Panessa06}, we estimate that this corresponds to a 2–10~keV flux limit of $\sim5 \times 10^{-15}$ erg cm$^{-2}$ s$^{-1}$. The expected H$\beta$/H$\alpha$ flux ratio for AGN has a wide range; as discussed in \cite{Mejia-Restrepo2022}, it can vary between 0.3 to 0.9, with an average of $\sim$0.5 for Seyfert 1 galaxies. Considering a worst-case scenario, we require objects in the redshift range 1.9 to 2.9 to have at least 3 times the required flux for the H$\alpha$ measurement. We apply these criteria to the objects in the \citet{Gilli07} 100~deg$^2$ catalog. 

To sum up, we require our object to be (i) bright enough to provide a reliable X-ray excess variance measurement in the 2-10 keV range, (ii) unobscured, (iii) bright enough to obtain a reliable redshift measurement using \textit{Euclid}, (iv) in the redshift range 0.88$<z<2.9$, so that the redshift could be provided using H$\alpha$ or H$\beta$.
We find a total of $\sim$4500 AGN that would satisfy the requirements for \textit{AXIS} (considering serendipitous sources over the mission lifetime, for 100 deg$^2$), and $\sim$3000 for the planned surveys of \textit{NewAthena} (for a total of 44 deg$^2$).

This sample size allows binning in redshift while maintaining adequate statistics per bin. Within a narrow redshift bin, the spread in luminosity distance is small, allowing us to average over the bin to reduce the scatter. Dividing the redshift range $0.84 < z < 2.9$ into 9 bins evenly spaced in log $z$, we find that each bin contains between 80 and 700 sources for the \textit{AXIS} sample, and between 40 and 500 sources for the \textit{NewAthena} one.  For reference, based on the fit presented in Section \ref{sec: distances}, we have $\delta \log(D_L)  \sim 0.5$ dex.
Assuming the uncertainty in flux is negligible, the uncertainty in the average $\log D_L$ for a bin with $N$ AGN is reduced by a factor of $\sqrt{N}$. This reduces the uncertainty on $\log(D_L)$ in each bin down to between 0.02 and 0.08~dex.\\
In Figure~\ref{fig:HD}, we present the resulting Hubble Diagrams for \textit{AXIS} (left) and \textit{NewAthena} (right). The AGN data points, colour-coded by whether \textit{Euclid} would observe H$\alpha$ or H$\beta$, are compared with Type~Ia supernovae from the Pantheon+ compilation. We also show predictions for the $\Lambda$CDM model and two alternative cosmologies, namely a $\Lambda$CDM with a different mass parameter ($\Omega_m = 0.1$), and a $w_0w_a$ model where dark matter is assumed to be varying with $w=w_0 = w_a(1+z)$. The divergence among models becomes significant only at $z \gtrsim 2$, highlighting the importance of high-redshift standard candles. While Type Ia supernovae are scarce beyond $z \sim 1.5$, AGN are more abundant in this regime, making them promising tools for future cosmological studies. As shown, a sample of thousands of objects would allow us to distinguish between cosmological models at high redshift. While the uncertainties increase at the highest redshifts, combining AGN with other cosmological probes in this range would allow discrimination among cosmological models.

Such increased statistics obtained by combining \textit{Euclid} with \textit{AXIS} and/or \textit{NewAthena} allow us to perform multiple additional tests. First, we can test if there are redshift variations in the relation. This is relevant for understanding the physics of variability, but also for validating the cosmological implementation. We can also further test the relation by dividing the sample into groups based on other quantities, such as luminosity or $\lambda_\mathrm{Edd}$, and examine whether the relation changes. 
Though encouraging, the results in Figure \ref{fig:HD} have some caveats. As we do not have a systematic study of the relation at higher redshift, it might be that the relation changes, the dispersion changes, or both. Furthermore, we assume that the intrinsic systematic scatter in the correlation is small and that the statistical scatter is dominant. This might not be the case, or might evolve with redshift as well. Hence, the presented results are a possible scenario only. More work is needed to better understand the relation, and be able to derive robust cosmological implications of such a promising Hubble diagram. 
The redshift for these objects might be provided by other surveys and/or telescopes in the future. In Appendix \ref{app: cosmo} we provide the Hubble Diagrams for the best case scenario in which all objects with an AXIS or \textit{NewAthena} measurement would have a redshift estimate. In that case, \textit{NewAthena} in particular would allow us to analyze the Hubble Diagram up to redshift $\sim$5.

Another important possible future synergy of the X-ray excess variance method is with the $L_X-L_{UV}$ relation in quasars \citep[e.g.,][]{RL19_nature, Trefoloni24, Lusso25}. This well-studied non-linear relation allows using quasars to measure cosmological distances over a wide redshift range ($z\sim0.7-4$); careful sample selection (removing quasars with dust reddening, X-ray absorption, etc.) reduces the observed dispersion of the $L_X-L_{UV}$ relation to $\sim$0.2 dex, and recent analyses indicate an intrinsic scatter of only $\leq0.1$ dex \citep[e.g.,][]{Lusso20, Sacchi22, Signorini24}. In the future, these two methods may be combined, each dominating in different redshift regimes. The $L_X-L_{UV}$ technique is most reliable at $z\geq$0.7 but becomes problematic at lower redshifts because rest-frame UV luminosity must be extrapolated and host-galaxy light can significantly contaminate the optical/UV flux. In contrast, X-ray excess variance is way less affected by host contamination, so the method can be applied to lower-$z$ AGN. Consequently, the two methodologies have the potential to operate synergistically, and the common redshift range could be used to double-check the calibration with SNIa. Notably, one contributor to the remaining $\sim$0.2 dex scatter in the $L_X-L_{UV}$ relation is the non-simultaneity of the measurements – X-ray variability means that a given UV state can correspond to a range of X-ray luminosities. This effect is especially pronounced in lower-luminosity quasars. Indeed, \cite{Signorini24} estimates that stochastic X-ray variability contributes on the order of $\delta\sim$0.08 dex to the dispersion of the $L_X-L_{UV}$ relation. By directly quantifying AGN X-ray variability (via excess variance) in individual sources, one could correct for or mitigate this variability-induced scatter in the $L_X-L_{UV}$ relation method, potentially tightening that relation further. In general, a deeper understanding of X-ray variability coming from future X-ray observatories will benefit both approaches: it can improve the precision of the $L_X-L_{UV}$ relation and refine the X-ray excess variance method. The combined use of these techniques would extend AGN “standard candle” coverage over a broad redshift range and allow critical cross-checks (through their overlap) to ensure consistent distance calibrations.

Finally, a possible extension of this work with future data regards the AGN monitoring that will be performed by the Vera C. Rubin Observatory Legacy Survey of Space and Time (LSST) \citep{LSST19}. Optical variability happens on longer time scales and is harder to characterise in terms of its excess variance, compared to X-ray variability. At the same time, the statistics that LSST will provide are unprecedented, with tens of millions of objects monitored over time \citep{DeCicco21}, so that the high statistics can still offer a promising path to use AGN variability to measure cosmological distances.

\section{Conclusions} 
\label{sec: conclusions}
In this work, we have investigated the relation between X-ray variability and key AGN physical properties using a sample of 134 Seyfert 1 galaxies from the BASS survey. By measuring excess variance on 20~ks light curves we reduced the fraction of upper limits from 30\% to 16\%, thereby improving the reliability of statistical analyses. Our main findings can be summarized as follows:

\begin{enumerate}
    \item We confirmed a strong anti-correlation between X-ray excess variance and both $M_{\rm BH}$ and different monochromatic luminosities. In contrast to earlier results, we found that the $\sigma^2_{\rm NXS} - M_{\rm BH}$ and $\sigma^2_{\rm NXS} - L$ relations have comparable scatter, and the residuals do not show significant trends with either parameter. However, when accounting for the intrinsic scatter associated with virial $M_{\rm BH}$ estimates, the data remain consistent with $M_{\rm BH}$ being the primary physical driver of the observed variability.

    \item We explored the dependence of variability on $\lambda_{\rm Edd}$, using refined values based on recent SED fitting \citep{Gupta24}. We found no significant correlation, contrary to predictions from the universal PSD scenario. A possible explanation is that the large uncertainties in virial mass estimates obscure the trend. When considering only the subset of AGN with reverberation mapping (RM) masses, we found an anti-correlation with $\lambda_{\rm Edd}$, but with low significance.

    \item We tested the application of the $\log(L_{2-10\,{\rm keV}}) - \log(\sigma^2_{\rm NXS})$ relation to estimate AGN intrinsic luminosities and hence distances. The increased sample size allowed us to reduce the scatter in the $\log L_{2-10\,{\rm keV}} - \log \sigma^2_{\rm NXS}$ relation to 0.63~dex, significantly lower than the 1.33~dex found by \citet{LaFranca14}. This translates to a dispersion of $\sim 0.32$~dex in $\log D_L$.

    \item We also tested whether including broad-line FWHM measurements, as proposed by \citet{LaFranca14}, could further reduce the dispersion. Contrary to their results, we found no significant improvement when using H$\alpha$ or H$\beta$. A moderate reduction in scatter was observed when using Pa$\alpha$ and Pa$\beta$, but the sample size for these lines remains small. 

    \item We simulated how future X-ray observatories such as \textit{AXIS} and \textit{NewAthena} could improve this methodology. These telescopes will enable precise measurements of excess variance for thousands of AGN. Considering the redshift values provided by \textit{Euclid}, we could extend the Hubble Diagram out to $z \sim 3$, potentially making X-ray variability a viable tool for cosmology.
\end{enumerate}

Further progress in this field will benefit from dedicated follow-up studies designed to refine both the physical calibration and the cosmological applicability of the X-ray variability–luminosity relation. Targeted reverberation-mapping campaigns of a representative subset of AGN in the BASS sample could provide more accurate black hole mass estimates, reducing one of the major sources of scatter and allowing a clearer assessment of the intrinsic dependence of variability on accretion rate. Improved constraints on $\lambda_{\rm Edd}$ could, in turn, reveal trends that might help standardise the relation further and understand the underlying physics better. Another possible future path is long-term X-ray monitoring for a statistically significant AGN sample. Extending the monitoring over months to years, we could measure the full PSD and determine its break timescale. As we know the break timescale correlates with the luminosity \citep[e.g.,][]{McHardy06}, that could provide a more constrained quantity to derive luminosities, and therefore distances. Together, these efforts will be key to advancing AGN variability  to a precision cosmological tool.\\

\begin{acknowledgements} 
We thank the referee for the useful comments and suggestions. MS, FR, SB acknowledge financial support from the Italian Ministry for University and Research, through the grant PNRR-M4C2-I1.1-PRIN 2022-PE9-SEAWIND: Super-Eddington Accretion: Wind, INflow and Disk-F53D23001250006-NextGenerationEU. AT acknowledges financial support from the Bando Ricerca Fondamentale INAF 2022 Large Grant “Toward an holistic view of the Titans: multi-band observations of $z>6$ QSOs powered by greedy supermassive black holes". MS also acknowledges support through the European Space Agency (ESA) Research Fellowship Programme in Space Science. DBS gratefully acknowledges support from NSF Grant 2407752. AJ acknowledges FONDECYT postdoctoral grant 323003. KO acknowledges support from the Korea Astronomy and Space Science Institute under the R\&D program (Project No. 2025-1-831-01), supervised by the Korea AeroSpace Administration, and the National Research Foundation of Korea (NRF) grant funded by the Korea government (MSIT) (RS-2025-00553982). BT acknowledges support from the European Research Council (ERC) under the European Union's Horizon 2020 research and innovation program (grant agreement number 950533). RS acknowledges funding from the CAS-ANID grant number CAS220016. ET and FEB acknowledge support from ANID CATA-BASAL program FB210003, and FONDECYT Regular 1241005 and 1250821. CR acknowledges support from SNSF Consolidator grant F01$-$13252, Fondecyt Regular grant 1230345, ANID BASAL project FB210003 and the China-Chile joint research fund. \textit{Softwares:} \textsc{TOPCAT} \citep{Taylor2005}, \textsc{SciPy} \citep{Virtanen2020}, \textsc{pandas} \citep{McKinney2010}, \textsc{Matplotlib}\citep{Hunter2007}.

\end{acknowledgements}

\bibliographystyle{aa} 
\bibliography{bibl}

\appendix

\section{Validation of the fitting method}
\label{app: method}
Our goal in this work is to fit linear relations in the logarithmic space, whether it be the excess variance versus black hole mass, luminosity, or luminosity combined with FWHM. Our analysis faces two main challenges: handling asymmetric uncertainties and incorporating upper limits.

First, both the dependent and independent variables have non-negligible uncertainties. To account for errors on both axes, we use the \texttt{lmfit} Python package, which accepts weights in both $x$ and $y$. However, our uncertainties are initially estimated in linear units. When converting to logarithmic space, the upper and lower uncertainties become asymmetric. For small uncertainties, averaging these errors to determine the weights might be acceptable; yet in our case—where the uncertainties in both excess variance and black hole masses (or a combination of luminosity and FWHM) are considerable—averaging introduces a bias that affects the final fit.

To overcome this, we proceed iteratively. We first average the upper and lower uncertainties and perform an initial fit. Using the best-fit line as a reference, we then determine for each data point whether it lies above or below the line and assign the corresponding upper or lower uncertainty. We repeat the fitting process with these adjusted uncertainties until the slope and intercept converge. Finally, we estimate the uncertainties on the fitted parameters via bootstrapping: for N=500 iterations, we sample each ($x,y$) value from its uncertainty distribution, perform the fit, and then derive the distributions of the slope and intercept. Tests with mock samples—constructed with the same uncertainty distribution as our real data—confirm that this method reliably recovers the input slope and intercept.

\begin{figure}
	\centering
\includegraphics[width=1\linewidth]{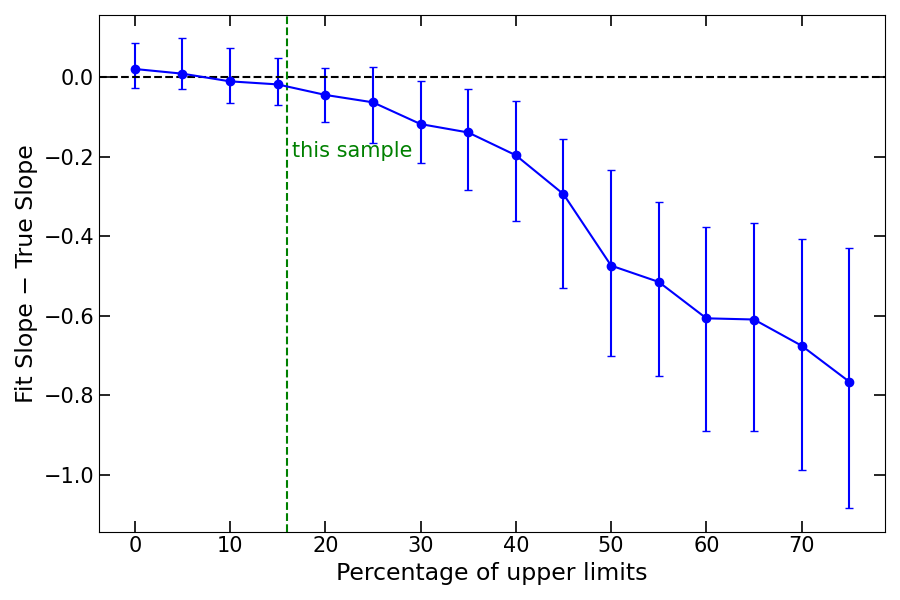}
\caption{Difference between the true slope chosen to build the mock sample and the one recovered with the fit procedure. Each point shows the mean and standard deviation obtained from 250 mock samples created with a given percentage of upper limits. The upper limit percentage in the BASS sample is 16\%, and we expect our fitting method to still be able to determine the correct slope.}
	\label{fig:mocksample_slopes}
\end{figure}

\begin{figure}
	\centering
\includegraphics[width=1\linewidth]{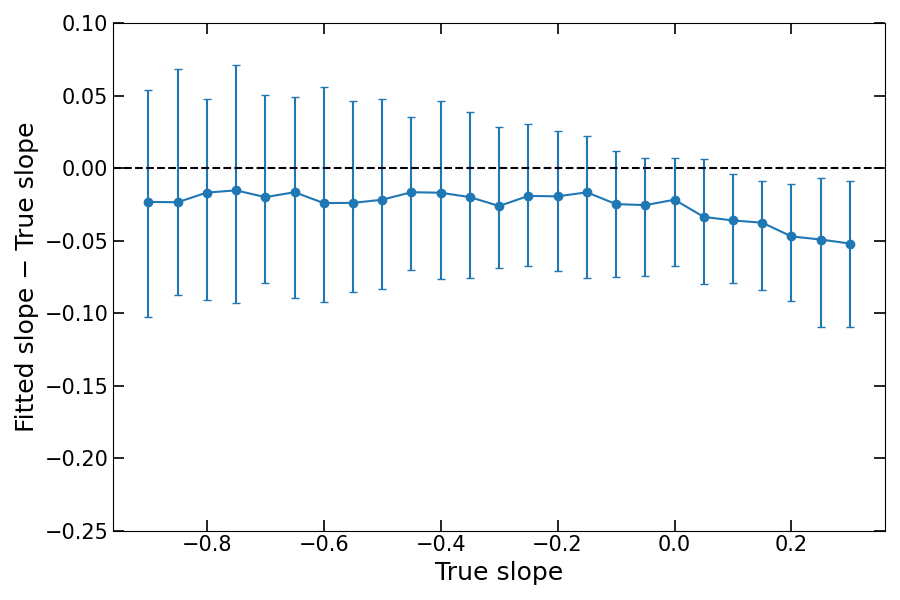}
\caption{Difference between the true slope chosen to build the mock sample and the one recovered with the fit procedure, assuming 15\% fraction of upper limits, and varying the value of the true slope. Each point shows the mean and standard deviation obtained from 250 mock samples created with a given slope.}
	\label{fig:mocksample_deltaslopeslope}
\end{figure}

Another issue is the presence of upper limits in the excess variance measurements, for which there is no overall consensus on the proper treatment. In our case, excess variance values are measured in linear units and then converted to logarithmic units for the fit. In linear space, we define an upper limit when the uncertainty equals the measured excess variance (i.e., when the value is consistent with zero). However, since the logarithm of zero is undefined, we set the minimal value to 10$^{-9}$ and derive the corresponding lower uncertainty in logarithmic space. This approach is not perfect; if the fraction of upper limits increases significantly, the fit may become biased. In particular, since the upper limits affect the $y$ variable predominantly at high $x$ values, they tend to bias the slope. We tested this by producing mock samples with an increasing fraction of upper limits, starting with an input slope of $s=-0.75$. For each fraction, we generate 250 mock samples and calculate the mean and standard deviation of the resulting slopes. We found that as long as the upper limit fraction remains below $\sim$20\%, the recovered slope is fully consistent with the assumed value. Between 20\% and 30\%, the value is still consistent with the input within 1$\sigma$, although systematically lower, and then becomes no more consistent when the upper limit ratio goes above 30\% (see Figure \ref{fig:mocksample_slopes}). Since our sample has an upper limit fraction of 16\%, we are confident that our method remains robust. We also tested whether the difference between the true slope and the fitted slope depends on the slope itself. In the case of a 15\% fraction of upper limits, we see that this is not the case, as shown in Figure~\ref{fig:mocksample_deltaslopeslope}. We tested on the range of slopes that we find in this work, between $\sim$-0.9 and $\sim$0.3. We see a small systematic shift of the fitted slope value for values above 0.10, but it is always consistent with the true value in a 1.2$\sigma$ level.

\section{Adding FWHM}
\label{app: fwhm}
\begin{table*}[ht]
\caption{Summary of the fitting results for the relation $\log(\sigma^2)$ vs. luminosities and the FWHM with the slope $\beta$ set to 4.}
\centering
\resizebox{\linewidth}{!}{%
\begin{tabular}{lcccccc}
\hline
Relation & $\alpha$ & $\gamma$ & $\delta_\mathrm{tot}$ (dex) & $\mathrm{err}_\mathrm{tot}$ (dex) & $\delta_\mathrm{int}$ (dex) & $N$ \\
\hline
$\log(\sigma^2)$ vs. $\log(L_{2-10 \, \mathrm{keV}}) + 4\log(\mathrm{FWHM}(H\alpha))$ & $-0.38 \pm 0.06$ & $14.90 \pm 0.09$ &  0.72 & 0.18 & 0.70 & 114 \\ %done
$\log(\sigma^2)$ vs. $\log(L_{6200 \, \text{\AA}}) + 4\log(\mathrm{FWHM}(H\alpha))$ & $-0.59 \pm 0.06$ & $24.63 \pm 0.07$ & 0.81 & 0.20 & 0.79 & 114 \\ %done 

$\log(\sigma^2)$ vs. $\log(L_{2-10 \, \mathrm{keV}}) + 4\log(\mathrm{FWHM}(H\beta))$ &$-0.33\pm0.05$ & $12.17\pm0.07$ & 0.72 & 0.20 & 0.69 & 87 \\ %done
$\log(\sigma^2)$ vs. $\log(L_{5100 \, \text{\AA}}) + 4\log(\mathrm{FWHM}(H\beta))$ & $-0.33\pm0.05$ & $12.58\pm0.07$ & 0.73 & 0.21 & 0.70 & 87 \\ % done

$\log(\sigma^2)$ vs. $\log(L_{2-10 \, \mathrm{keV}}) + 4\log(\mathrm{FWHM}(Pa\alpha))$ & $-0.55\pm0.06$ & $22.58\pm0.08$ & 0.69 & 0.27 & 0.63 & 40\\ % done
$\log(\sigma^2)$ vs. $\log(L_{2-10 \, \mathrm{keV}}) + 4\log(\mathrm{FWHM}(Pa\beta))$ & $-0.66\pm0.08$& $28.14\pm0.10$ &0.70 & 0.28 & 0.64 & 51\\ %done

\hline
\end{tabular}%
}

\label{tab:results_withFWHM_4}
\end{table*}

\begin{table*}[ht]
\caption{Summary of the fitting results for the relation $\log(\sigma^2)$ vs. luminosities and the FWHM with the slope $\beta$ set to 2.}
\centering
\resizebox{\linewidth}{!}{%
\begin{tabular}{lcccccc}
\hline
Relation & $\alpha$ & $\gamma$ & $\delta_\mathrm{tot}$ (dex) & $\mathrm{err}_\mathrm{tot}$ (dex) & $\delta_\mathrm{int}$ (dex) & $N$ \\
\hline
$\log(\sigma^2)$ vs. $\log(L_{2-10 \, \mathrm{keV}}) + 2  \log(\mathrm{FWHM}(H\alpha))$ & $-0.50 \pm 0.07$ & $20.18 \pm 0.08$ & 0.71 & 0.18 & 0.69 & 114 \\ %done
$\log(\sigma^2)$ vs. $\log(L_{6200 \, \text{\AA}}) + 2 \log(\mathrm{FWHM}(H\alpha))$ & $-0.75 \pm 0.06$ & $32.33 \pm 0.07$ & 0.83 & 0.18 & 0.81 & 114 \\ %done

$\log(\sigma^2)$ vs. $\log(L_{2-10 \, \mathrm{keV}}) + 2 \log(\mathrm{FWHM}(H\beta))$ & $-0.49\pm0.06$ & $18.39\pm0.07$ & 0.70 & 0.20 & 0.67 & 87 \\ 
$\log(\sigma^2)$ vs. $\log(L_{5100 \, \text{\AA}}) + 2 \log(\mathrm{FWHM}(H\beta))$  & $-0.46\pm0.06$ & $17.09\pm0.06$ & 0.74 & 0.20 & 0.71 & 87 \\ 

$\log(\sigma^2)$ vs. $\log(L_{2-10 \, \mathrm{keV}}) + 2\log(\mathrm{FWHM}(Pa\alpha))$ & $-0.75\pm0.08$ & $32.10\pm0.08$ & 0.65 & 0.30 & 0.58 & 40\\ %done

$\log(\sigma^2)$ vs. $\log(L_{2-10 \, \mathrm{keV}}) + 2\log(\mathrm{FWHM}(Pa\beta))$ & $-0.84\pm0.09$ & $36.14\pm0.10$ & 0.67 & 0.28 & 0.61 & 51\\ %done

\hline
\end{tabular}%
}

\label{tab:results_withFWHM_2}
\end{table*}

In \cite{LaFranca14}, the authors reduced the dispersion in the relation between the excess variance and the luminosity by including the measurements of the FWHM of broad lines. The reasoning starts from the assumption that the primary driver of the observed relation is the $M_\mathrm{BH}$, with the luminosity following as a secondary effect, given its dependence on the mass itself. If this is indeed the case, it is possible to combine the variance-mass and mass-luminosity relations to derive:
\begin{equation}
    \log(\sigma^2) = \alpha [\log(L) + 4 \log(\mathrm{FWHM})] + \gamma.
    \label{eq:with_fw}
\end{equation}
By calibrating this relation, one can estimate an AGN's intrinsic luminosity from its X-ray excess variance and broad-line FWHM. This, in turn, allows its distance to be determined. \citet{LaFranca14} incorporated FWHM from H$\beta$ and Pa$\beta$ into their analysis, observing significant reductions in the observed dispersion with respect to the case of luminosity alone versus luminosity and FWHM together. In particular, they found: from 1.78 to 1.12~dex for the $L_{5100 \,\mathrm{\AA}}$-H$\beta$ combination, from 1.36 to 1.06~dex for the $L_{2-10\,\mathrm{keV}}$-H$\beta$ combination, and from 1.33 to 0.71~dex for the $L_{2-20\,\mathrm{keV}}$-Pa$\beta$ combination. However, their sample sizes were relatively small, comprising 31, 38, and 18 objects, respectively, for the three combinations. 

In the present work, we do not observe a smaller dispersion when fitting the excess variance against black hole mass than against luminosity. Therefore, we do not expect the inclusion of the FWHM to change the observed dispersion. Still, for completeness and to compare with the results in \cite{LaFranca14}, we performed the analysis presented here. Our dataset includes 88 objects with H$\beta$ measurements, more than double the sample size used by \citet{LaFranca14}, and 49 objects with Pa$\beta$ measurements, nearly triple their sample size. Additionally, we test this method with H$\alpha$ and Pa$\alpha$, for which we have 115 and 39 objects, respectively. 
To test Equation \ref{eq:with_fw}, we analyzed different combinations of monochromatic luminosities and FWHM. For each of the four broad emission lines tested, we used the X-ray luminosity in the 2–10~keV band as the luminosity variable. Additionally, for H$\beta$, we tested the monochromatic luminosity at 5100~\AA, while for H$\alpha$, we tested the luminosity at 6200~\AA. Our goal was to test these optical/UV luminosities close to the broad lines as they may correlate more tightly with $M_\mathrm{BH}$ in the context of the virial equation. Results are shown in Table \ref{tab:results_withFWHM_4}, in Figures C.1-C.5.

We compared our results for the 2–10~keV luminosity with the baseline total dispersion ($\delta_\mathrm{tot} = 0.73$~dex) and intrinsic dispersion ($\delta_\mathrm{int} = 0.70$~dex) obtained previously. Using H$\alpha$ or H$\beta$ as the FWHM indicator, we observed a slight improvement, with $\delta_\mathrm{tot}$ reducing to 0.72~dex and $\delta_\mathrm{int}$ remaining the same (for H$\alpha$) or decreasing slightly (for H$\beta$). Thus, the inclusion of H$\alpha$ or H$\beta$ has a limited impact. We also tested whether replacing the X-ray luminosity with the optical monochromatic luminosities at 5100~\AA\ (H$\beta$) and 6200~\AA\ (H$\alpha$) could further reduce the dispersion. However, as shown in Table \ref{tab:results_withFWHM_4}, this approach resulted in higher dispersion values (see also the figures in Appendix \ref{app: plots}). While monochromatic luminosities near the emission line should, theoretically, better trace the $M_\mathrm{BH}$ in the virial equation, the stronger correlation between X-ray luminosity and excess variance may be more significant in practice. This could be attributed to the fact that X-ray and optical/UV observations are not simultaneous. Indeed, \citet{LaFranca14} also found smaller dispersion values when using X-ray luminosities rather than optical ones.  When analyzing the Paschen lines, we observed a more substantial reduction in dispersion. For Pa$\alpha$ and Pa$\beta$, the total dispersion was reduced to $\delta_\mathrm{tot} = 0.69$ and 0.70~dex, respectively, with intrinsic dispersions of $\delta_\mathrm{int} = 0.63$ and 0.64~dex. These tighter relations are notable, despite the smaller sample sizes for the Paschen lines. Compared to \citet{LaFranca14}, we observed lower dispersion values for all combinations of luminosities and lines common to that study. Doubling or tripling the sample size in our analysis appears to significantly improve the results, suggesting that further increases could yield even tighter correlations. However, contrary to \citet{LaFranca14}, we do not observe a significant reduction in dispersion when using H$\beta$ or H$\alpha$. 

Our analysis assumes that the virial factor $f$ in the virial equation is constant. However, this is a simplification, as the virial factor likely varies across objects due to differences in the geometry and kinematics of the BLR. 
Recently, the direct modelling of the BLR emission in reverberation mapping campaigns has been proposed as a way to directly measure the virial factor $f$ for single sources. However, such analyses are time-consuming and require a very high SNR to be performed \citep{Villafana24}. Indeed, only $\sim$40 AGN have undergone such modeling \citep{Villafana23}. Still, even if we cannot directly measure $f$ for all the objects in our sample, we can explore the effect of the dependence of the virial factor on other observable quantities. We tested whether the virial factor $f$ depends on the FWHM of broad lines, as suggested by previous studies. A dependence of $f \propto \mathrm{FWHM}^{-1}$ has been suggested \citep[e.g.,][]{Mejia-Restrepo18, Yu19}, based on the inclination dependence of the BLR. If this dependence holds, the slope associated with the FWHM term in Equation \ref{eq:with_fw} would change from 4 to 2. Using this relation, we repeated our analysis, with the results presented in Table \ref{tab:results_withFWHM_2}. For the H$\beta$ and H$\alpha$ lines combined with $L_{2-10 \,\mathrm{keV}}$, we observed reduced dispersions compared to not including the FWHM term. The improvement was even more pronounced for the Paschen lines, with the combination of Pa$\alpha$ and $L_{2-10 \,\mathrm{keV}}$ achieving the lowest intrinsic dispersion, $\delta_\mathrm{int} = 0.58$~dex. We also tested different values for the dependence of $f$ on the FWHM, with $f\sim$FWHM$^{-\eta}$ with $\eta$ in the range from 0 to 2, and we still found that values around $\eta\sim$1 provide the smallest observed dispersion. These results suggest that the virial factor may indeed depend on the FWHM, potentially due to inclination effects. Incorporating BLR modeling and objects with RM data could provide a more robust sample for calibration. However, RM samples may still be biased, as they often include AGN with smaller $M_\mathrm{BH}$, which exhibit stronger variability, making monitoring campaigns more feasible. Addressing these biases and incorporating inclination effects into the analysis represent directions for future work.

\section{Excess variance versus X-ray Eddington ratio}
\label{app:Xrayedd}
\begin{figure*}
    \centering
    \includegraphics[width=0.48\textwidth]{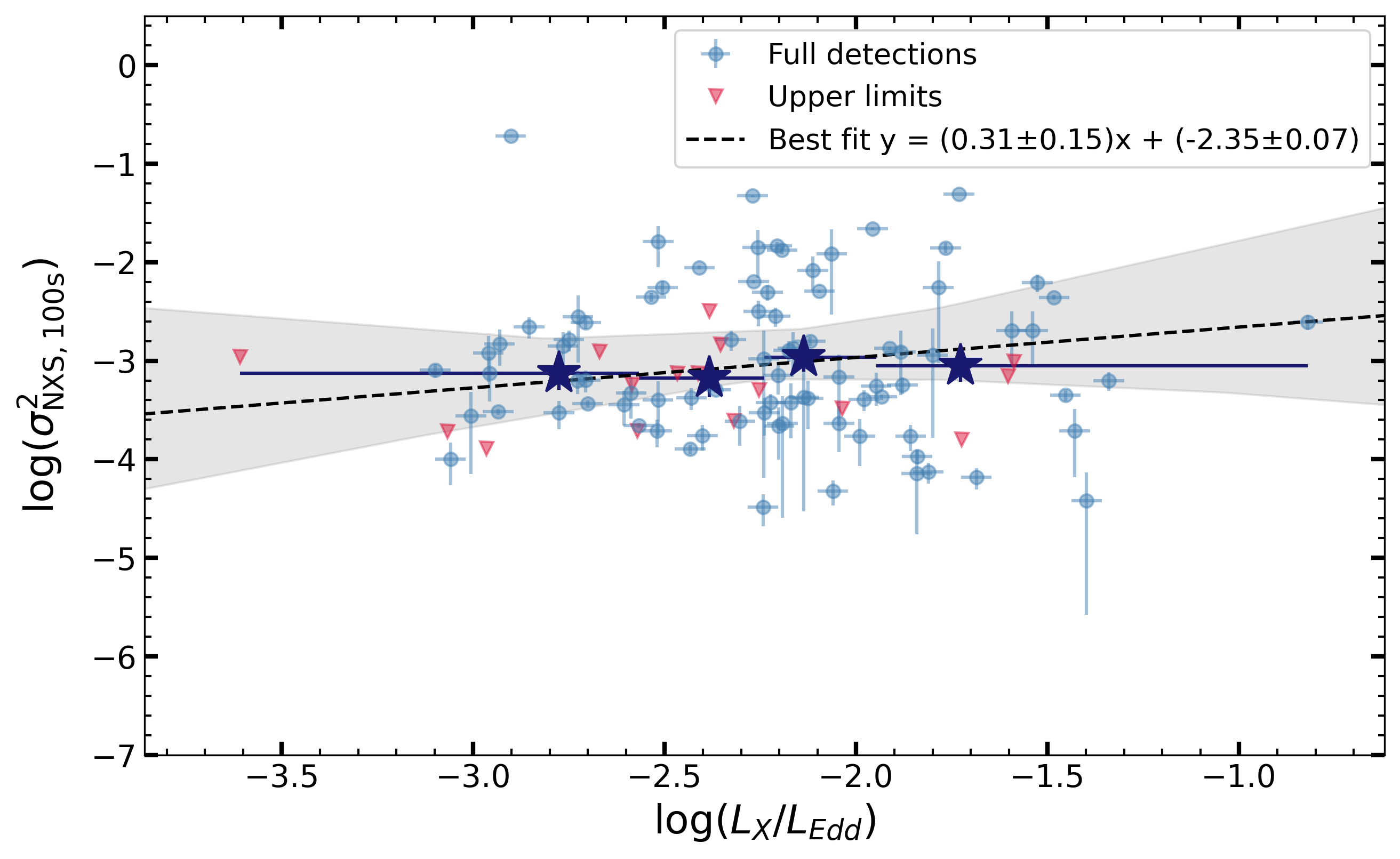}
    \includegraphics[width=0.48\textwidth]{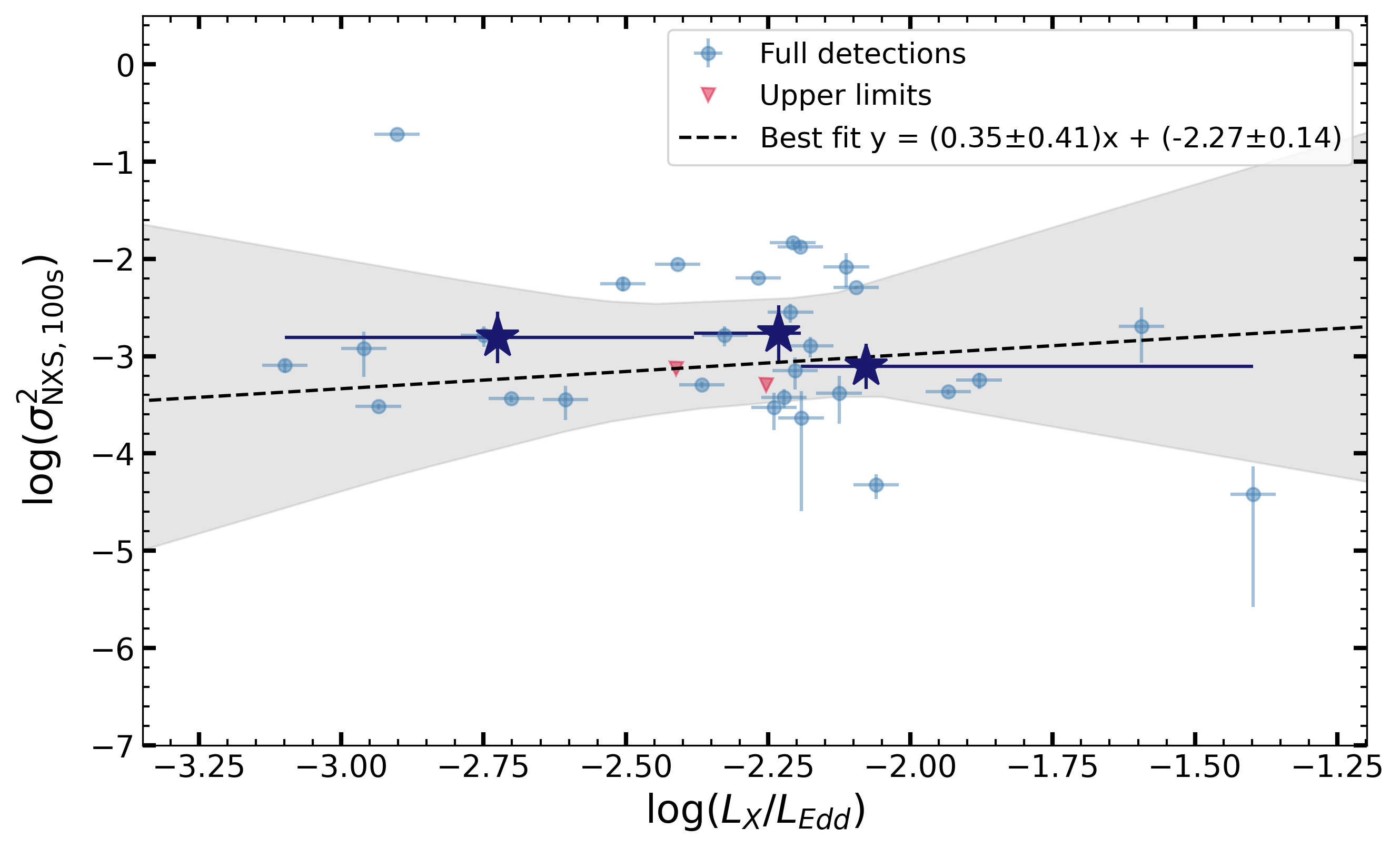}
    \caption{Relation between the X-ray excess variance and the X-ray Eddington ratio, $\lambda_{\rm X} \equiv L_{2-10\,{\rm keV}}/L_{\rm Edd}$. 
    \textit{Left}: full sample; \textit{Right}: RM-only subsample. Cyan points show detections; red triangles show upper limits. Dark-blue stars indicate medians in $\lambda_{\rm X}$ bins (shown for visualisation only). The black dashed lines show the best-fit relations; the grey shaded areas show the 1$\sigma$ confidence intervals. Slopes and intercepts are indicated in each panel.}
    \label{fig:exvar_lxledd_all}
    \label{fig:exvar_lxledd_RM}
\end{figure*}

We tested whether the short-timescale X-ray excess variance correlates with the X-ray Eddington ratio, $\lambda_{\rm X} \equiv \frac{L_{2-10\,{\rm keV}}}{L_{\rm Edd}}$, $L_{\rm Edd} = 1.26\times10^{38}\,\left(\frac{M_{\rm BH}}{M_\odot}\right)\,{\rm erg\,s^{-1}}$. We fit a log-linear relation of the form
$\log\!\left(\sigma^2_{\rm NXS}\right) = a\,\log(\lambda_{\rm X}) + b$,
using the iterative procedure described in Sect.~3, and we report results for both the full sample and the RM-only subsample.
For the full sample we find a weak positive slope $a_{\rm all} = +0.31 \pm 0.15$,
which differs from zero at the $\sim$2.1$\sigma$ level. For the RM-only subsample the slope is $a_{\rm RM} = +0.35 \pm 0.41$, fully consistent with no correlation. The corresponding fits and 1$\sigma$ confidence bands are shown in Fig.~\ref{fig:exvar_lxledd_all} left and right panels. 
Interpreting $\sigma^2_{\rm NXS}$ versus $\lambda_{\rm X}$ is non-trivial because $\log L_{\rm Edd}$ is effectively proportional to $\log M_{\rm BH}$, and $L_X$ and $M_{\rm BH}$ are themselves strongly correlated in our sample. Consequently, $\lambda_{\rm X}$ introduces covariance between variables already entering the $\sigma^2$–$M_{\rm BH}$ and $\sigma^2$–$L_X$ relations, making the slope sensitive to that.
Given these caveats and the marginal significance, we refrain from drawing a physical conclusion from the weak positive trend. In the main text (Sect.~4.4), we therefore adopt the analysis based on bolometric Eddington ratios derived from SED fitting.

\section{Additional figures}
\label{app: plots}
Here we provide the plots for the fit of $\log \sigma^2_{NXS} = \alpha \log(L) + \beta FWHM + \gamma$, when using the monochromatic luminosity at 5100\AA{} and 6200\AA{} when fitting using the FWHM of the H$\beta$ and H$\alpha$ lines, respectively. On one hand, using a monochromatic continuum closer to the emitted line should be a more accurate choice for the virial equation for the BH mass measurement. However, we observe a worse overall performance in terms of observed dispersion when using optical monochromatic continuum luminosities compared to 2-10 keV monochromatic luminosities. This could be because, as the X-ray and optical observations are not simultaneous, the X-ray luminosity is a better proxy as it more strongly correlates with the X-ray excess variance. 

\begin{figure*}[htbp]
    \centering
    \includegraphics[width=0.49\textwidth]{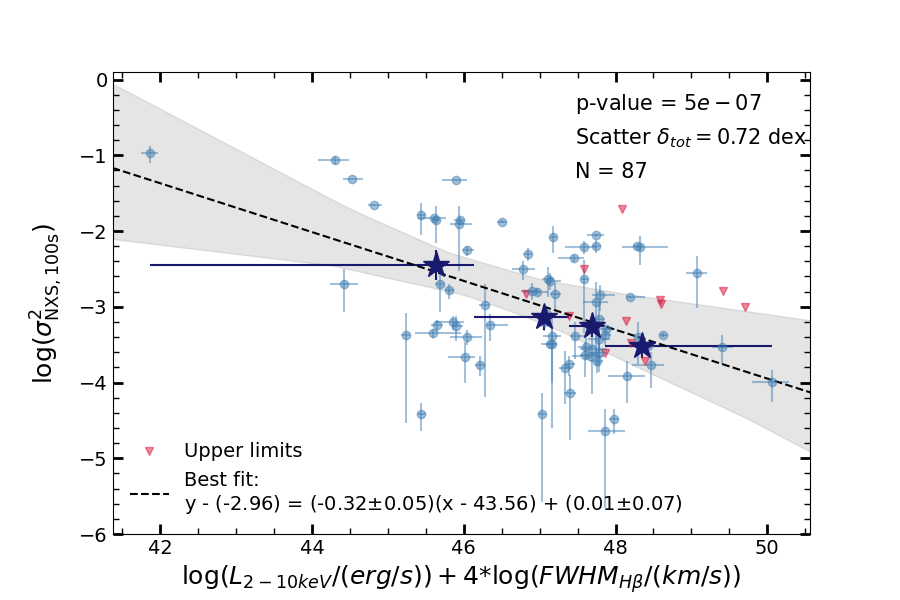}
    \hfill
    \includegraphics[width=0.49\textwidth]{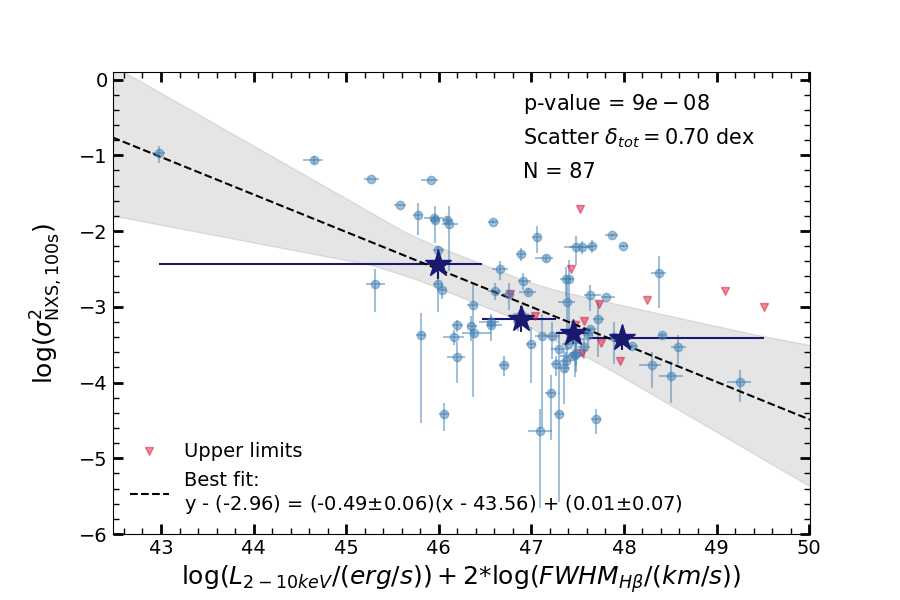}
    \caption{Relation between the X-ray excess variance and the combination of monochromatic luminosity and broad line FWHM for the H$\beta$ line, $\sigma^2_{NXS} = \alpha (\log(L) + \beta FWHM) + \gamma$. N=87; the legend is the same as the previous Figures. Left: $L$ is the 2-10 keV luminosity, $b$=4, as predicted from the virial equation. Right: $L$ is the 2-10 keV luminosity, $b$=2. When $b$=2, we observe lower dispersion.}
    \label{fig:exvar_Hb_L210}
\end{figure*}

\begin{figure*}[htbp]
    \centering

\end{figure*}

\begin{figure*}[htbp]
    \centering
    \includegraphics[width=0.49\textwidth]{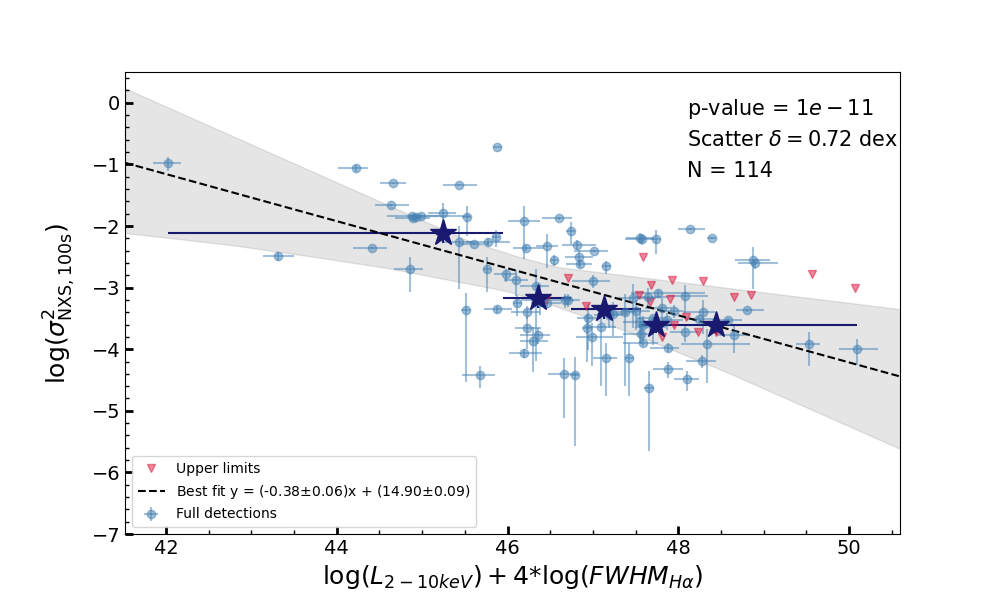}
    \hfill
    \includegraphics[width=0.49\textwidth]{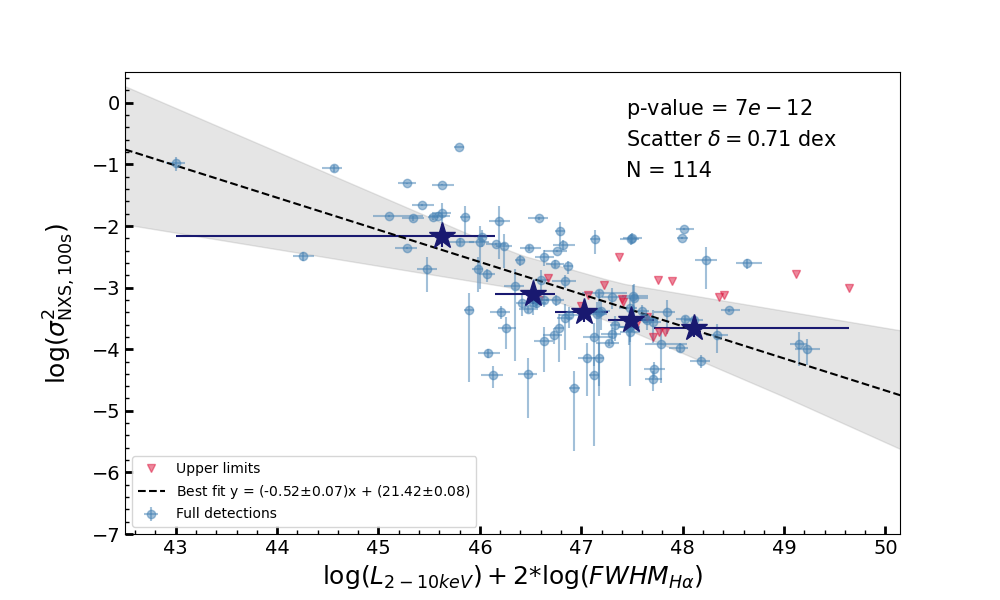}
    \caption{The same as Figure \ref{fig:exvar_Hb_L210} but for the H$\alpha$ broad line. N=114. Again, we see that the dispersion slightly lowers going from $b$=4 to $b$=2.}
    \label{fig:exvar_Ha_L210}
\end{figure*}

\begin{figure*}[htbp]
    \centering
    \includegraphics[width=0.49\textwidth]{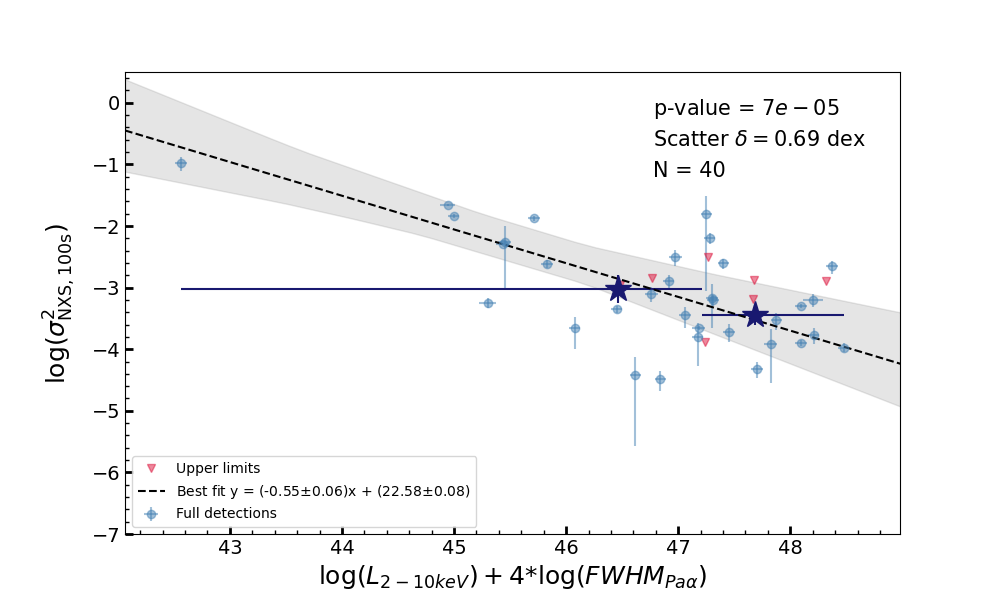}
    \includegraphics[width=0.49\textwidth]{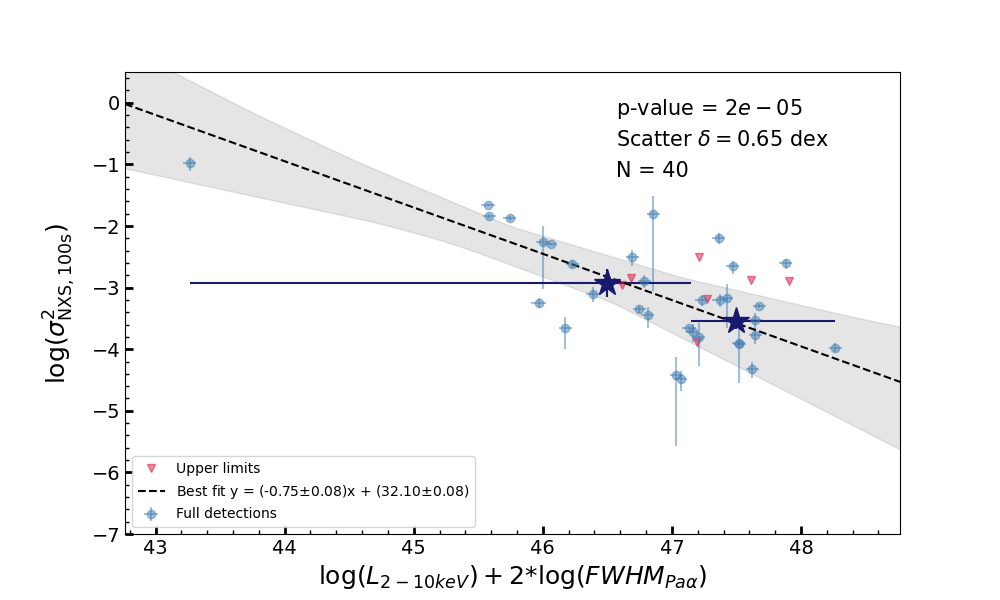}
    \caption{The same as Figure \ref{fig:exvar_Hb_L210} but using the Pa$\alpha$ broad line FWHM. N=40. We observe a reduction in dispersion when going from $b$=4 to $b$=2.}
    \label{fig:exvar_L_FWHM_paschenA}
\end{figure*}

\begin{figure*}[htbp]
    \centering
    \includegraphics[width=0.49\textwidth] {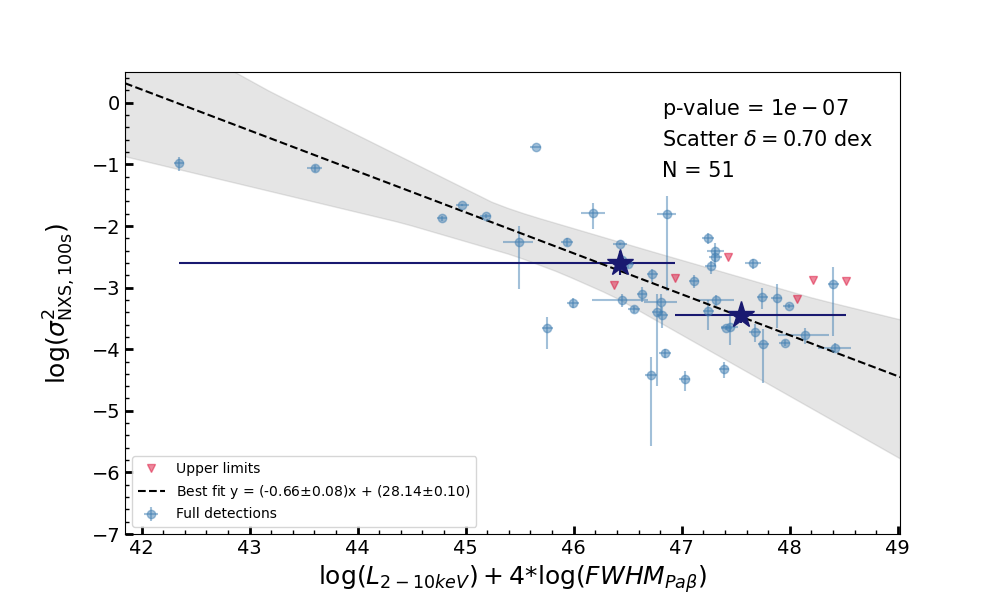}
    \includegraphics[width=0.49\textwidth]{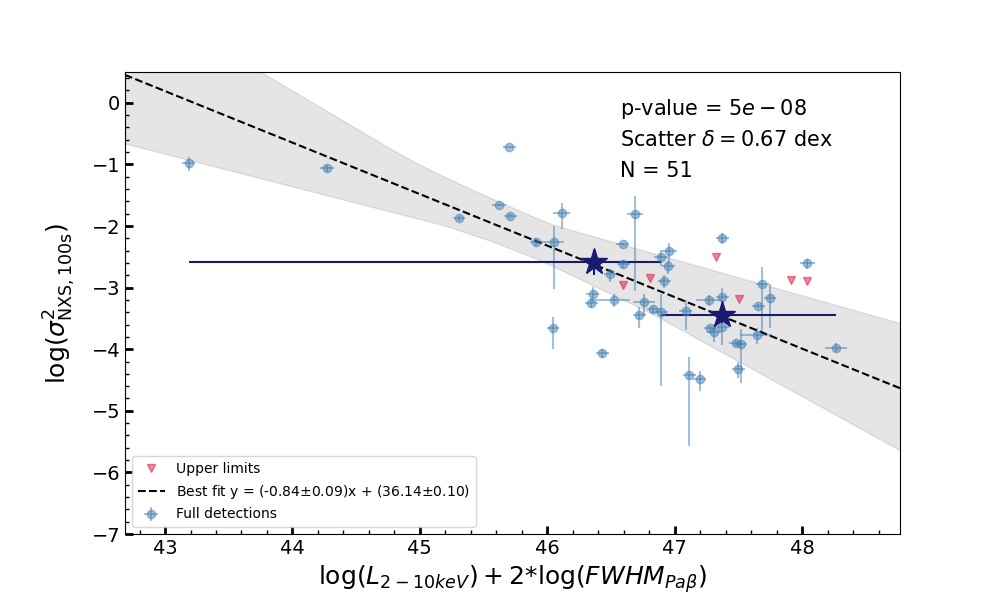}
    \caption{The same as Figure \ref{fig:exvar_Hb_L210} but using the Pa$\beta$ broad line FWHM. N=56. We observe a reduction in dispersion when going from $b$=4 to $b$=2.}
    \label{fig:exvar_L_FWHM_paschenB}
\end{figure*}

\begin{figure*}[htbp]
    \centering
    \includegraphics[width=0.49\textwidth]{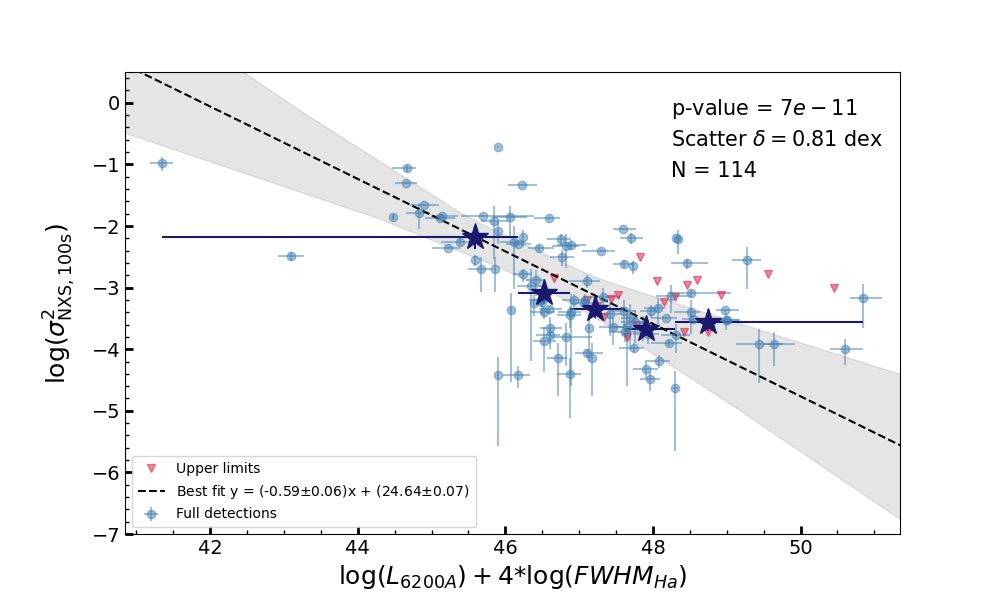}
    \hfill
    \includegraphics[width=0.49\textwidth]{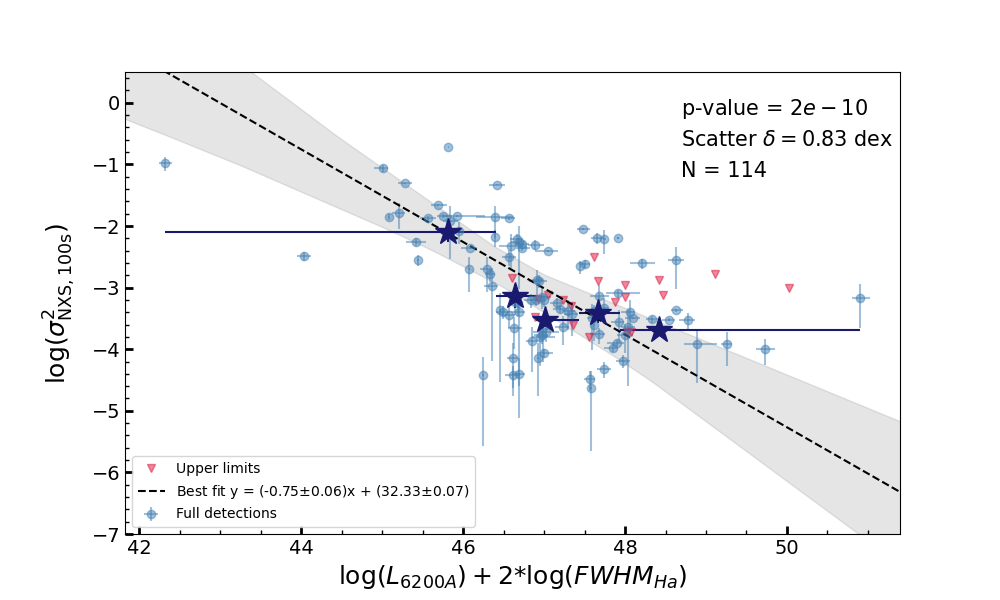}
    \caption{The same as Figure \ref{fig:exvar_Hb_L210} but for the H$\alpha$ broad line, using the 6200 \AA{} monochromatic luminosity. N=114. }
    \label{fig:exvar_Ha_L5100}
\end{figure*}

\section{Additional cosmological plots}
\label{app: cosmo}
In Section \ref{sec: future}, we performed the simulation considering the objects that would have X-ray observations with AXIS and/or \textit{NewAthena}, and a spectroscopic redshift value obtained using Euclid. It could be that a redshift estimate would be obtained with other probes. Here, in this Appendix, we show the Hubble Diagram for the 'best case scenario' in which all objects with an X-ray excess variance measurement obtained by AXIS and/or \textit{NewAthena} would have a spectroscopic redshift. In particular, the right panel in Figure D.1 shows how with the deep surveys planned for \textit{NewAthena} we would be able to reach up to redshift $\sim$5, testing the expansion rate of the Universe when it was less than a billion years old.

\begin{figure*}[htbp]
    \centering
    \includegraphics[width=0.49\textwidth]{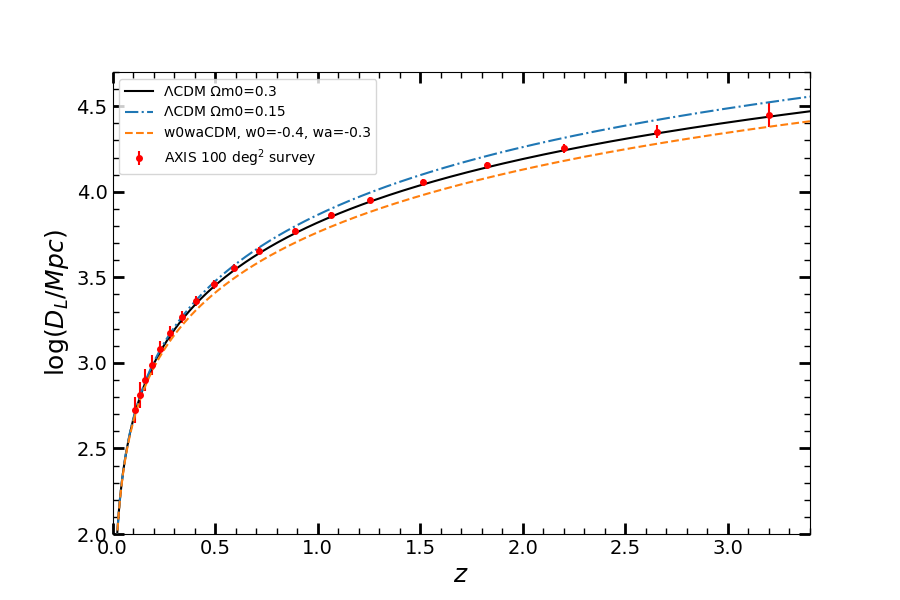}
    \hfill
    \includegraphics[width=0.49\textwidth]{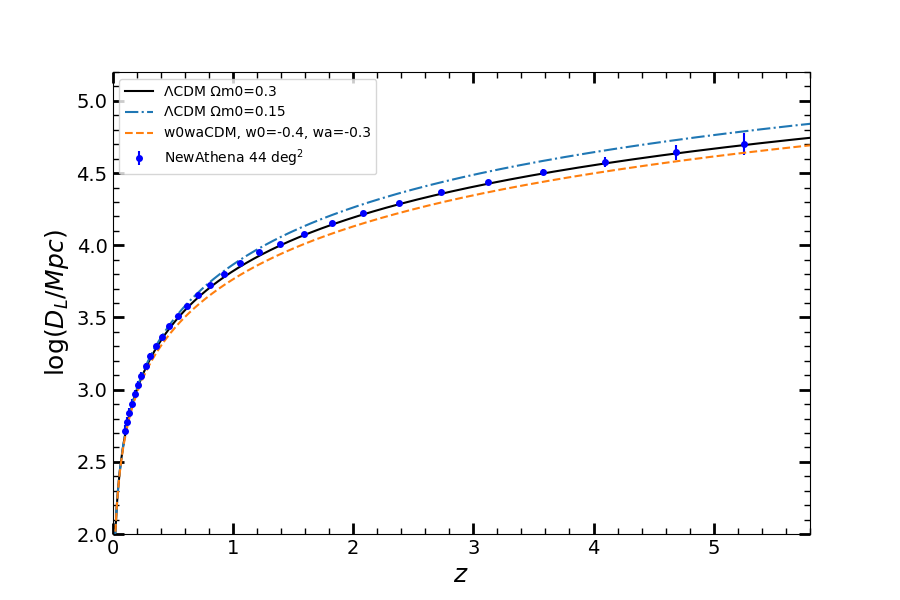}
    \caption{left: simulated Hubble Diagrams obtained by using the relation between X-ray excess variance and 2-10 keV luminosity to derive the luminosity distance $D_L$. The diagram is generated for a sample of objects expected to be observed by AXIS, assuming they will all have a spectroscopic redshift measurement. Right: same simulation but for the \textit{NewAthena} telescope.\\ }
    \label{fig:HD_app}
\end{figure*}

\end{document}